\documentclass[iop]{emulateapj-rtx4}
\usepackage{natbib}
\usepackage{color}
\usepackage{graphicx}
\begin{document}

\newcommand{\Ka}{K$\alpha$}
\newcommand{\Kb}{K$\beta$}
\newcommand{\NH}{$N_{\rm H}$}
\newcommand{\Msun}{$M_{\odot}$}
\newcommand{\chisq}{$\chi ^2$/dof}
\newcommand{\lumi}{ergs~s$^{-1}$}
\newcommand{\flux}{ergs~cm$^{-2}$~s$^{-1}$}
\newcommand{\sbn}{ergs~cm$^{-2}$~s$^{-1}$~arcmin$^{-2}$}
\newcommand{\kms}{km s$^{-1}$}

\newcommand{\redpen}[1]{{\bf[\textcolor{red}{#1}]}}
\newcommand{\bluepen}[1]{{\textcolor{blue}{#1}}}
　　　

\shorttitle{Fe-Rich Knot in Tycho's SNR}
\shortauthors{Yamaguchi et al.}


\title{The Origin of the Iron-Rich Knot in Tycho's Supernova Remnant}

\author{
Hiroya Yamaguchi\altaffilmark{1,2},  
John P.\ Hughes\altaffilmark{3},
Carles Badenes\altaffilmark{4},
Eduardo Bravo\altaffilmark{5},  
Ivo R.\ Seitenzahl\altaffilmark{6},\\
H{\'e}ctor Mart{\'i}nez-Rodr{\'i}guez\altaffilmark{4}, 
Sangwook Park\altaffilmark{7},  
Robert Petre\altaffilmark{1}
}
\email{hiroya.yamaguchi@nasa.gov}

\altaffiltext{1}{NASA Goddard Space Flight Center, Code 662, Greenbelt, MD 20771, USA}
\altaffiltext{2}{Department of Astronomy, University of Maryland, College Park, MD 20742, USA}
\altaffiltext{3}{Department of Physics and Astronomy, Rutgers University, 136 Frelinghuysen Road, 
	Piscataway, NJ 08854, USA}
\altaffiltext{4}{Department of Physics and Astronomy and Pittsburgh Particle Physics, 
	Astrophysics and Cosmology Center (PITT PACC), University of Pittsburgh, 
	3941 O'Hara St, Pittsburgh, PA 15260, USA}
\altaffiltext{5}{E.T.S.\ Arquitectura del Vall{\`e}s, Universitat Polit{\`e}cnica de Catalunya, 
	Carrer Pere Serra 1-15, 08173 Sant Cugat del Vall{\`e}s, Spain}
\altaffiltext{6}{Research School of Astronomy and Astrophysics, 
	The Australian National University, Cotter Road, Weston Creek, ACT, 2611, Australia}
\altaffiltext{7}{Department of Physics, University of Texas at Arlington, 
	Box 19059, Arlington, TX 76019, USA}

\begin{abstract}

X-ray observations of supernova remnants (SNRs) allow us to investigate the chemical 
inhomogeneity of ejecta, offering unique insight into the nucleosynthesis in supernova explosions. 
Here we present detailed imaging and spectroscopic studies of the ``Fe knot'' located along the eastern 
rim of the Type Ia SNR {\it Tycho} (SN\,1572) using {\it Suzaku} and {\it Chandra} long-exposure data. 
Surprisingly, the {\it Suzaku} spectrum of this knot shows no emission from Cr, Mn, or Ni, 
which is unusual for the Fe-rich regions in this SNR. 
Within the framework of the canonical delayed-detonation models for SN Ia, the observed 
mass ratios $M_{\rm Cr}/M_{\rm Fe} < 0.023$, $M_{\rm Mn}/M_{\rm Fe} < 0.012$, and 
$M_{\rm Ni}/M_{\rm Fe} < 0.029$ (at 90\% confidence) can only be achieved for a peak 
temperature of (5.3--5.7)\,$\times 10^9$\,K and a neutron excess of $\lesssim 2.0 \times 10^{-3}$. 
These constraints rule out the deep, dense core of a Chandrasekhar-mass white dwarf as 
the origin of the Fe knot, and favors either incomplete Si burning or the $\alpha$-rich 
freeze-out regime, probably close to their boundary. An explosive He burning regime 
is a possible alternative, although this hypothesis is in conflict with the main properties 
of this SNR.

\end{abstract}


\section{Introduction}

Type Ia supernovae (SNe Ia), widely believed to originate from thermonuclear explosions 
of white dwarfs, play an important role in the chemical evolution of the universe, 
as they release a large amount of heavy elements synthesized during the explosion. 
SNe Ia are also crucial for the study of cosmology, owing to their use as distance indicators. 
Nevertheless, many of their fundamental aspects still remain poorly understood \cite[e.g.,][]{Maoz14}.

X-ray observations of supernova remnants (SNRs) provide unique insight into 
the nucleosynthesis that had taken place in their progenitor, because they allow us to investigate the 
amount and distribution of heavy elements via spatially-resolved spectral analysis \citep[e.g.,][]{Hwang12,Park07}. 
{\it Tycho}'s SNR, the remnant of SN\,1572, is an ideal object in that sense, 
since its spatial structure is well resolved owing to its proximity and moderate angular 
size \citep[e.g.,][]{Decourchelle01,Warren05}.  {\it Tycho}'s SNR is thought to be the result of 
a typical SN Ia explosion with a normal brightness that synthesized $\sim$\,0.7\Msun\ of $^{56}$Ni, 
based on spectroscopy of the optical light echo \citep{Krause08} and an X-ray study combined with 
hydrodynamical simulations \citep{Badenes06}.

\begin{figure*}[t]
  \begin{center}
  	\includegraphics[width=12cm]{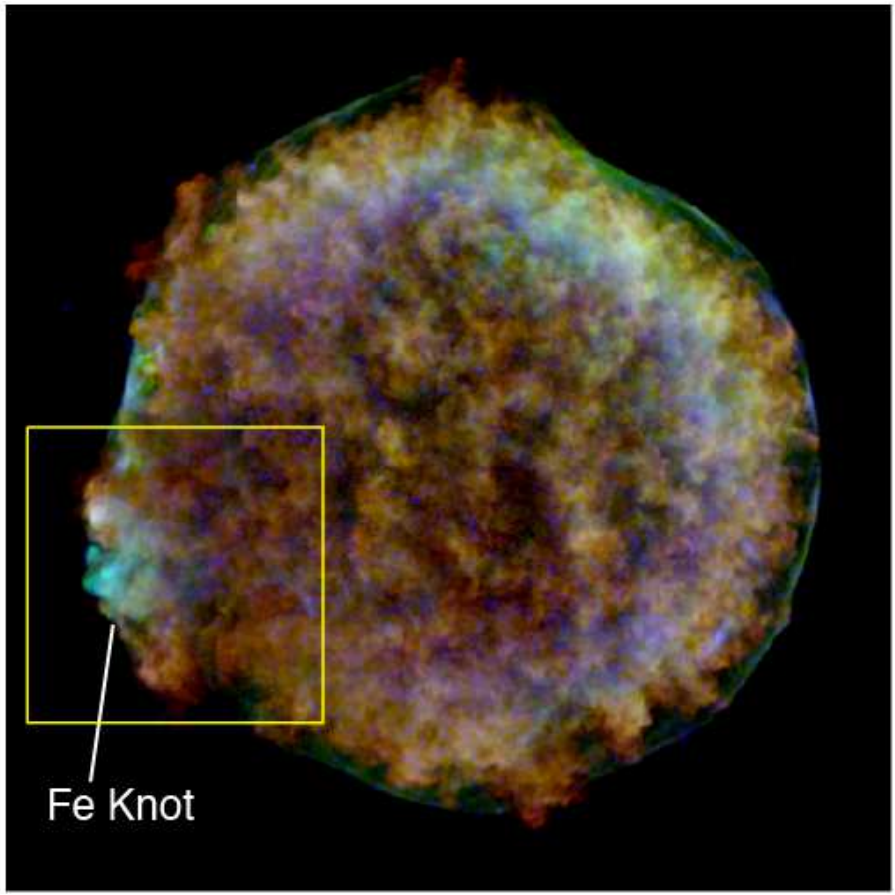}
	\vspace{1mm}
\caption{
Three-color image of {\it Tycho}'s SNR from the {\it Chandra} deep observation. 
Red, green, blue are emission from the Si K (1.8--1.92\,keV), Fe L (0.8--1.25\,keV), and Fe K (6.35--6.6\,keV) 
bands, respectively. The yellow box indicates the region shown in Figure\,\ref{fig:image-cxo}.
  \label{fig:image}}
  \end{center}
\end{figure*}

The chemical inhomogeneity in {\it Tycho}'s SNR has been studied extensively. 
{\it ASCA} observations revealed a stratified elemental composition throughout most of the SNR, 
with Fe interior to the intermediate-mass elements (IMEs: e.g., Si, S, Ar, Ca) \citep{Hwang97,Hwang98}. 
This stratification is consistent with predictions from modern numerical simulations \citep[e.g.,][]{Seitenzahl13a} 
as well as actual observations of an SN Ia explosion \citep[e.g.,][]{Mazzali07,Tanaka10}. 
On the other hand, the eastern region of the SNR exhibits somewhat unusual morphological features: 
several `clumps' overrunning the forward shock have been spatially resolved 
\citep[][see also Figure\,\ref{fig:image}]{Vancura95}. 
Given the enhanced abundance of Fe confirmed by {\it XMM-Newton} \citep{Decourchelle01}, 
one of the clumps (hereafter ``Fe knot'') is likely to have originated from a relatively hot region in 
the exploding progenitor, where a large amount of $^{56}$Ni was generated.
Interestingly, the location of the Fe knot coincides with an apparent `gap' in the reverse shock 
structure identified by {\it Chandra} \citep{Warren05}. 
This coincidence suggests an association between the mechanism that created the Fe knot and 
the bulk dynamics of the explosion. However, to date there has been no extensive investigation of 
the nature of this interesting knot based on an X-ray spectroscopy.

More recently, {\it Suzaku} has opened a new window into the physics of both SNe Ia 
and their remnants with its high sensitivity for weak lines, such as \Ka\ emission of the secondary 
Fe-peak elements (e.g., Cr, Mn, Ni, which are normally synthesized together with Fe) and 
\Kb\ fluorescence of Fe \citep[e.g.,][]{Tamagawa09,Yang13,Yasumi14}. 
For instance, the abundances of the secondary Fe-peak elements provide a probe 
of the neutron excess in the progenitor, due to either metallicity effects \citep{Badenes08b,Park13} 
or electron capture reactions during the SN explosion \citep{Yamaguchi15}.  
The centroid energy and intensity of the Fe \Kb\ fluorescence, on the other hand, constrain the physical conditions 
in the non-equilibrium plasma \citep{Yamaguchi14a}, which in turn enables accurate abundance measurements 
to be made. In this paper, we perform detailed imaging spectroscopy of {\it Tycho}'s Fe knot 
utilizing both the superior sensitivity of {\it Suzaku} and the excellent angular resolution of {\it Chandra}, 
and place strong constraints on its plasma state and origin.

Throughout this paper, the distance to the SNR is assumed to be 3\,kpc \citep[e.g.,][]{Ruiz04,Tian11}, 
but our main results and conclusions are not affected by its exact value 
\citep[see][for its systematic uncertainty from various literatures]{Hayato10}. 
The spectral analysis is all performed with the XSPEC software \citep{Arnaud96}.
The errors quoted in the text and table represent the 90\% confidence level, 
and the error bars given in the spectra represent 1$\sigma$ confidence.

\section{Observational Results}

We analyzed archival data of {\it Tycho}'s SNR obtained using the {\it Suzaku} X-ray Imaging Spectrometer (XIS) 
and the {\it Chandra} Advanced CCD Imaging Spectrometer (ACIS), the same datasets as used by 
\cite{Yamaguchi14a} and \cite{Eriksen11}, respectively. 
The details of the observations are summarized in Table\,\ref{tab:obs}. 
We reprocessed the data in accordance with the standard procedures using the latest calibration database, 
obtaining the total effective exposures given also in Table\,\ref{tab:obs}.  Figure\,\ref{fig:image} shows 
an ACIS high-resolution image of {\it Tycho}'s SNR, where the Fe knot at the eastern rim is indicated.

\begin{table}[t]
\begin{center}
\caption{Summary of the Observations.
  \label{tab:obs}}
  \begin{tabular}{lcc}
\hline \hline
Missions & {\it Suzaku} & {\it Chandra} \\
Instruments & XIS0 \& 3 & ACIS-I \\
Cycle & 3 (LP$^{*}$) & 10 (LP$^{*}$) \\
Observation ID & 5030850[1,2]0 & 1009[3--7], 1090[2--4,6] \\
Observation Date & 2008 Aug 4--12 & 2009 Apr 11--May 3  \\
Exposure Time (ks) & 416 & 734 \\ 
\hline
\end{tabular}
\tablecomments{
$^{*}$Large Program.
}
\end{center}
\end{table}

\begin{figure*}[t]
  \begin{center}
  	\includegraphics[width=16.6cm]{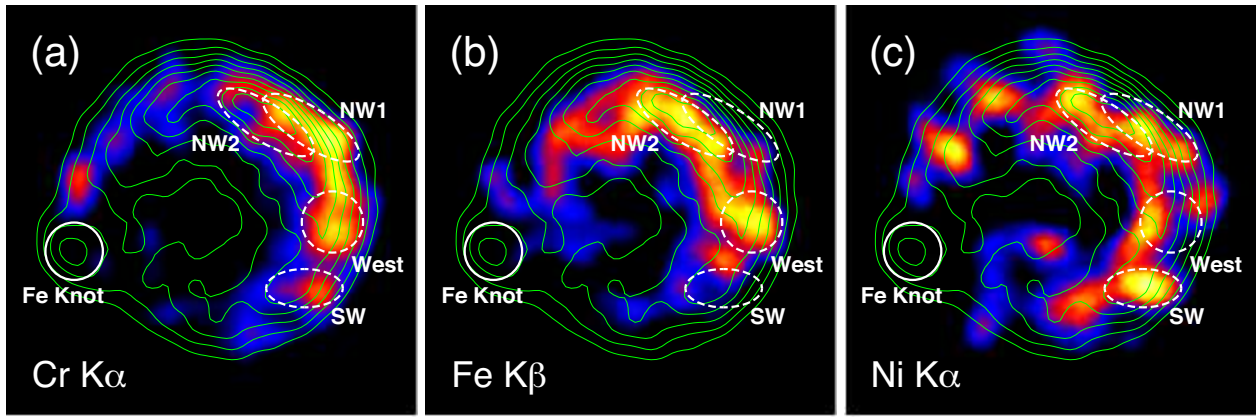}
	\vspace{1mm}
\caption{
 {\it Suzaku}/XIS smoothed images of {\it Tycho}'s SNR in the 5.38--5.58\,keV (a: Cr \Ka), 
 7.00--7.20\,keV (b: Fe \Kb), and 7.38--7.58\,keV (c: Ni \Ka) bands 
 where a 6.34--6.53\,keV (Fe \Ka) image is overplotted in contours. 
 The continuum flux estimated using a 7.7--9.0\,keV image is subtracted from the raw images. 
 The white solid circle is where an XIS spectrum of the Fe knot (Figure\,\ref{fig:spec-knot}) is extracted. 
 We also extract spectra from the other regions enclosed by the dashed circles or ellipses for comparison, 
 which are shown in Figure\,\ref{fig:spec-all}.
  \label{fig:image-suzaku}}
  \end{center}
\end{figure*}

\subsection{Suzaku Narrow Band Images}

Figure\,\ref{fig:image-suzaku} shows {\it Suzaku}/XIS images of the SNR at the energies of 
(a) Cr \Ka, (b) Fe \Kb, and (c) Ni \Ka\ emission, after subtracting the nonthermal continuum 
flux estimated using the procedure described in \cite{Yamaguchi14a}. We use only the data 
from the front-illuminated CCDs (XIS0 and 3) and merge them to improve the photon statistics.
The green contours overlaid on the images are taken from the Fe \Ka\ band, where the Fe knot 
is clearly seen at the east rim even with the lower angular resolution of the XIS images. 
No counterpart of the Fe knot is apparent in any of the other energy bands 
(i.e., Cr \Ka, Fe \Kb, and Ni \Ka). 
Although the primary topic of this paper is the Fe knot, we here briefly comment on 
other interesting features revealed by Figure\,\ref{fig:image-suzaku}. 
There is a distinct difference in the distribution of Cr and Fe (with the former at a larger radius) 
along the western side. This separation supports the idea suggested by 
\cite{Badenes08b} that the shocked Cr originates from explosive Si burning, 
whereas Fe comes from a mixture of Si burning and nuclear statistical equilibrium (NSE) burning. 
The even smaller peak radius of the Fe \Kb\ emission at the bright northwest (NW) region is 
due to the extremely low ionization of the ejecta immediately behind the shock \citep{Yamaguchi14a}. 
A second peak in the Fe \Kb\ image is found along the western rim, where the lowest ambient 
density is reported \citep{Williams13}. 
Although the detailed morphology of the Ni \Ka\ emission can provide important clues to SN Ia explosion 
physics (as we discuss is \S3), the photon statistics in Figure\,\ref{fig:image-suzaku}(c) are too poor to 
allow us to conclude anything about the actual Ni distribution in {\it Tycho}'s SNR.
More detailed study of these features is left for future work.

\subsection{Spectrum Extraction and Comparison}

\begin{figure}[t]
  \begin{center}
	\includegraphics[width=8cm]{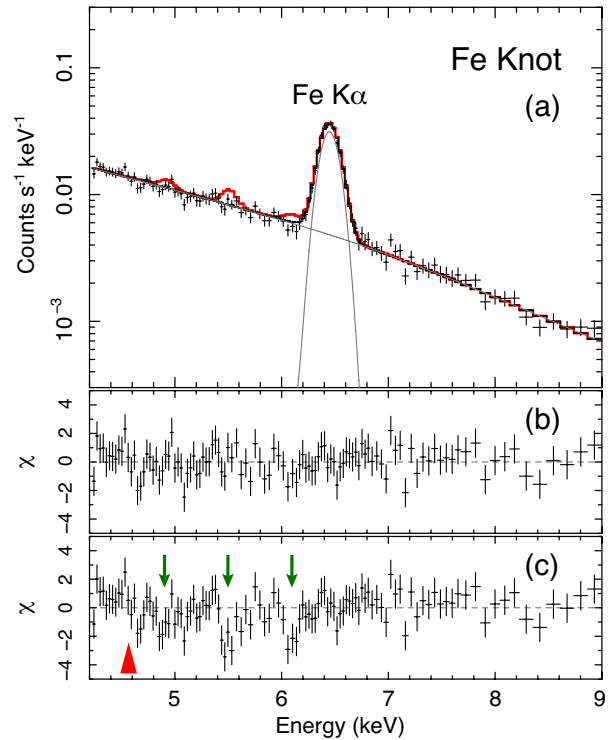}	
	\vspace{1mm}
\caption{(a) An XIS spectrum of the Fe knot in 4.2--9.0\,keV. Besides the strong Fe \Ka\ line, 
no significant line emission is detected. The red line shows the model reported by \cite{Miceli15} 
where detection of Ti (at $\sim$4.9\,keV), Cr, and Mn \Ka\ lines was claimed. \ 
(b) Residuals from our best-fit model consisting of only one Gaussian for the Fe \Ka\  
and a power-low for the continuum component. 
(c) Residuals from the model of \cite{Miceli15}. Dips are seen at the energies of 
the claimed lines (green arrows). The Ti \Ka\ centroid expected for the ionization timescale 
of the Fe knot ($n_e t \sim 1.2 \times 10^{10}$\,cm$^{-3}$\,s) is indicated by the red triangle in panel(c). 
See \S2.4 for more details.  
  \label{fig:spec-knot}}
  \end{center}
\end{figure}

\begin{figure*}[t]
  \begin{center}
	\includegraphics[width=15.8cm]{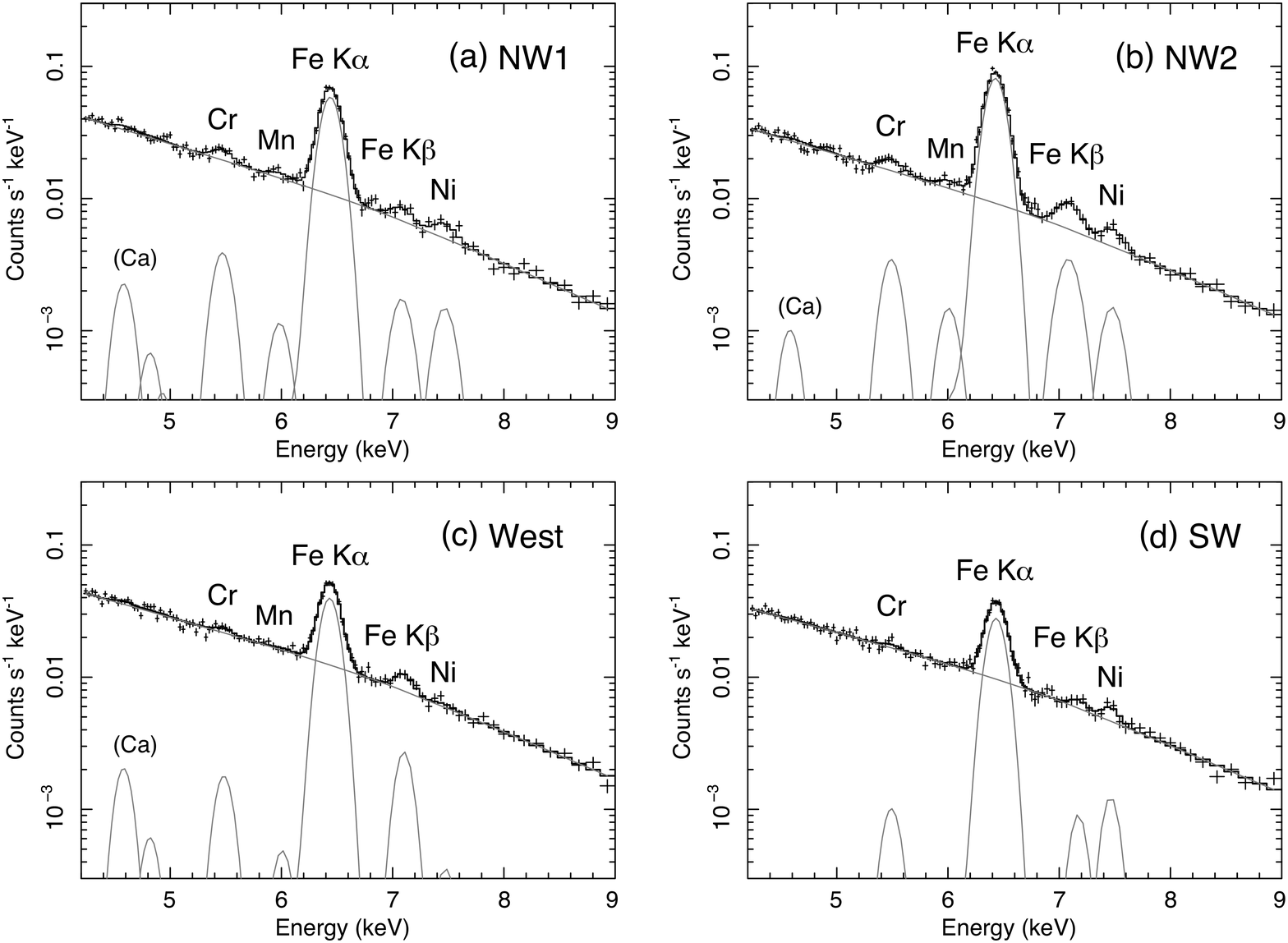}	
	\vspace{1mm}
\caption{XIS spectra of the (a) NW1, (b) NW2, (c) West and (d) SW regions given in Figure\,\ref{fig:image-suzaku}. 
The phenomenological model components (a power law and Gaussians) applied to determine the line parameters 
are shown as well.  
  \label{fig:spec-all}}
  \end{center}
\end{figure*}

For more quantitative studies, we extract an XIS spectrum of the Fe knot from the region 
in Figure\,\ref{fig:image-suzaku} marked by a circle at (R.A., Dec.)$_{\rm J2000}$ 
= (0h\,25m\,52.5s, +64$^\circ$07$'$10.$\!''$0) with a diameter of 1.$\!'$5. Since this value 
is smaller than the half-power diameter (HPD) of the {\it Suzaku} X-Ray Telescope 
\cite[XRT: $\sim$\,2.$\!'$0;][]{Serlemitsos07}, 
we study the effect of photon scattering using the {\tt xissim} task with the {\it Chandra} image 
at the Fe-K band as an input. We find that about 35\% of photons originating from 
the Fe knot are detected in the circular region (i.e., the remaining 65\% fell into the outer regions), 
but this detector region is still dominated by the photons from the Fe knot itself.
A spectrum of the non-X-ray background (NXB) for the same detector region is generated using 
the XIS night-Earth database and subtracted from the source spectrum. 
The resulting spectrum in the 4.2--9.0\,keV band is shown in Figure\,\ref{fig:spec-knot}(a). 
We find that the continuum level of the NXB is $\sim 4 \times 10^{-5}$\,counts\,s$^{-1}$\,keV$^{-1}$, 
well below the source spectrum. Therefore, a high signal-to-noise ratio is achieved in this region. 
We ignore the contribution of the cosmic X-ray background (CXB), since its flux is less than 
1\% of the source even at $\sim$\,9\,keV where the estimated CXB/source flux ratio would be highest. 
Although \Ka\ emission lines from the secondary Fe-peak elements, as well as Fe \Kb\ fluorescence, 
fall in this energy band (4.2--9.0\,keV), we see no evidence for them. 
For comparison, we also extract spectra from other regions (dashed circles or ellipses 
given in Figure\,\ref{fig:image-suzaku}) in Figure\,\ref{fig:spec-all}. 
The regions are chosen to contain bright spots of either Cr \Ka, Fe \Kb, or Ni \Ka\ emission.
Despite the significantly higher continuum level in these spectra than in the Fe knot region, 
they clearly show weak lines associated with these metal species. 
This indicates that the absence of these lines in the Fe knot spectrum is not due to 
insufficient energy resolution or photon statistics but indeed is due to their low line fluxes.

We fit all the spectra (Figure\,\ref{fig:spec-knot} and \ref{fig:spec-all}) with ad hoc 
Gaussian models to constrain the line centroids and flux ratios relative to Fe \Ka. 
The line widths of the weak or undetected lines are linked to that of the Fe \Ka\ line.
The results are given in Table\,\ref{tab:fit-suzaku}, where the flux upper limit for 
the undetected lines is determined using the procedures described in \S2.3 and \S2.4. 
The flux ratios are generally in good agreement with the imaging analysis; for instance, 
the NW1 region (brightest in the Cr \Ka\ image) has the highest Cr/Fe ratio. 
Interestingly, the Fe \Ka\ centroid energy of the Fe knot is highest among the analyzed regions, 
indicating that the Fe ejecta in this SNR are most highly ionized in the Fe knot. 
During the spectral fitting performed above, we model the continuum component with a power law, 
assuming absorption by the foreground interstellar medium (ISM) by a hydrogen column density 
of $7 \times 10^{21}$\,cm$^{-2}$ \citep{Cassam07} and standard ISM abundances \citep{Wilms00}. 
We also take into account the contribution of K$\beta$ and K$\gamma$ emission of He-like Ca by fixing 
their centroids and flux ratios at the theoretical values, although this does not affect the results we obtain.

\begin{table}[t]
\begin{center}
\caption{Statistics and best-fit parameters for the Suzaku spectra.
  \label{tab:fit-suzaku}}
  \tiny
    \begin{tabular}{lccccc}
\hline \hline
~ & Fe Knot & NW1 & NW2 & West & SW \\
\hline
\multicolumn{6}{c}{Photon Counts} \\
\hline
4.2--9.0\,keV$^{a}$ & 14390 & 32906 & 31104 & 33422 & 25372    \\
Fe \Ka$^{b}$ & 3050 & 5612 & 7881 & 3598 & 2557  \\
\hline
\multicolumn{6}{c}{Centroids [eV]} \\
\hline
Cr \Ka & (5480--5490)$^{c}$  & $5474_{-29}^{+27}$ & $5497_{-34}^{+29}$ & $5487 \pm 53$ & $5496_{-86}^{+88}$ \\
Mn \Ka & (5952--5960)$^{c}$ & $5989_{-78}^{+105}$ & $6012 \pm 65$& 6012 & 6012 \\
Fe \Ka & $6453_{-4}^{+3}$ & $6445 \pm 3$ & $6433_{-3}^{+2}$ & $6438_{-3}^{+4}$ & $6437 \pm 4$  \\
Fe \Kb & (7100--7400) & $7087_{-40}^{+46}$ & $7091 \pm 23$ & $7107_{-27}^{+29}$ & $7176_{-60}^{+58}$ \\
Ni \Ka & (7509--7514)$^{c}$ & $7473_{-50}^{+47}$ & $7482_{-39}^{+38}$ & 7482 & $7467_{-35}^{+40}$  \\
\hline
\multicolumn{6}{c}{Flux Ratios [\%]} \\
\hline
Cr/Fe (\Ka) & $<$\,2.5$^{d}$ & $5.6 \pm 1.6$ & $3.6_{-1.0}^{+1.1}$ & $3.8 \pm 2.4$ & $3.1 \pm 2.9$  \\
Mn/Fe (\Ka) & $<$\,1.0$^{d}$ & $1.8 \pm 1.4$ & $1.7 \pm 0.9$ & $1.1_{-1.1}^{+2.1}$ &  $0.1_{-0.1}^{+2.5}$  \\
Fe \Kb/\Ka & $<$\,2.3$^{d}$ & $3.7 \pm 1.3$ & $5.7_{-1.2}^{+1.4}$ & $7.6_{-2.2}^{+2.6}$ & $3.6 \pm 2.4$   \\
Ni/Fe (\Ka) & $<$\,2.6$^{d}$ & $3.8 \pm 1.4$ & $2.7 \pm 0.9$ & $1.3_{-1.3}^{+2.1}$ & $5.6 \pm 2.5$   \\
\hline
\end{tabular}
\tablecomments{
$^{a}$Data counts. $^{b}$Model predicted counts for the Gaussian component. 
$^{c}$Ranges predicted from the Fe \Ka\ centroid (see \S2.3 and \S2.4). 
$^{d}$The upper limits are at the 90\% confidence limit.
}
\end{center}
\end{table}

\subsection{Plasma Diagnostics for the Fe Knot}

\begin{figure}[t]
  \begin{center}
	\includegraphics[width=8.7cm]{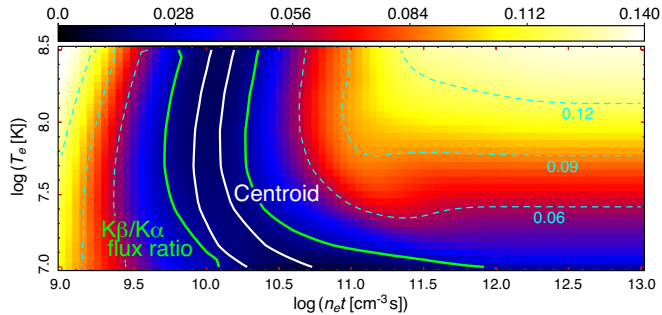}	
\caption{
  Theoretical predictions for the Fe \Kb/\Ka\ emissivity ratio as a function of the ionization 
  timescale ($n_e t$: horizontal axis) and electron temperature ($T_e$: vertical axis). 
  The regions constrained by the observed flux ratio and Fe \Ka\ centroid are indicated by 
  the green and white curves, respectively. To derive the latter, Figure\,3(c) of \cite{Yamaguchi15} 
  is used with account of the systematic uncertainty in the detector gain 
  \citep[0.1\% of the mean energy;][]{Ozawa09a}.  
  Note that the emissivity ratio is lowest around the constrained ionization timescale 
  ($\sim 1 \times 10^{10}$\,cm$^{-3}$\,s), because the dominant Fe ions at this plasma condition
  (Fe$^{16+}$ and Fe$^{17+}$) still have many $2p$ electrons but no $3p$ electron 
  that is responsible for the \Kb\ fluorescence. 
   \label{fig:atomic}}
  \end{center}
\end{figure}

The Fe \Ka\ centroid of the Fe knot ($6453_{-4}^{+3}$\,eV) corresponds to   
an average charge number of $\left< z_{\rm Fe} \right> = 17$ \citep{Yamaguchi14a}. 
Also sensitive to the Fe charge number is the \Kb/\Ka\ flux ratio, because the fluorescence yields 
of these lines depend on the number of bound electrons in the $2p$ and $3p$ shells. 
We find that this ratio for the Fe knot does not exceed 0.023 (at the 90\% confidence level) 
for any \Kb\ centroid energy between 7.1\,keV to 7.4\,keV. 
Figure\,\ref{fig:atomic} shows the theoretically-predicted Fe \Kb/\Ka\ emissivity ratio 
as a function of the ionization timescale (horizontal axis) and electron temperature (vertical axis), 
calculated using the latest {\it AtomDB} database\footnote{http://www.atomdb.org} 
\citep[see also][]{Yamaguchi15}, where the ionization timescale $n_e t$ is the product of 
the electron density and the elapsed time since the gas was shock heated. The regions constrained by 
the measured flux ratio and the \Ka\ centroid are fully consistent with each other (Figure\,\ref{fig:atomic}). 
This consistency indicates that the Fe K-shell spectrum of this knot is well characterized by a non-equilibrium 
ionization (NEI) plasma model with a {\it single} ionization timescale of $\sim 1.2 \times 10^{10}$\,cm$^{-3}$\,s, 
supporting the idea that the Fe knot is indeed an isolated ejecta clump, in contrast to the Fe ejecta in 
the main SNR shell where a broad range of ionization states has been measured \citep{Yamaguchi14a}.

\begin{figure}[t]
  \begin{center}
	\includegraphics[width=7.8cm]{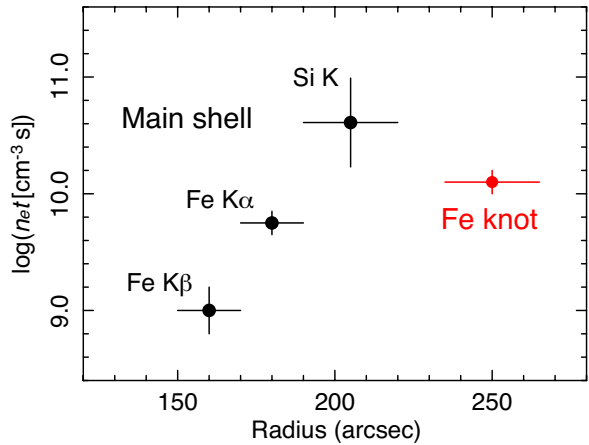}	
\caption{
	Relationship between the ionization timescale and the distance from the SNR center  
	for the Fe knot (red) and the bulk emission from the main SNR shell (black), clearly showing 
	that the Fe knot is dynamically distinct from the bulk of the ejecta. 
	The SNR center determined from 
	previous {\it Chandra} observations \citep{Warren05} is assumed.  
   \label{fig:r-nt}}
  \end{center}
\end{figure}

The different thermodynamic evolution of the Fe knot is illuminated more clearly in Figure\,\ref{fig:r-nt}, 
where its angular radius and ionization timescale are compared with those of the main SNR shell 
(i.e., bright NW region). We choose Si K, Fe \Ka, and Fe \Kb\ as the characteristic emission from 
the bulk of the ejecta, and plot (with the black points) their radial extent and ionization timescale 
reported in previous work \citep{Warren05,Badenes07,Yamaguchi14a}. 
The figure indicates that the exterior ejecta tend to be more highly ionized at the main SNR shell, 
as naturally expected from the reverse shock dynamics \citep[e.g.,][]{Badenes03}, 
while the Fe knot (the red point) is distinctly inconsistent with the trend.

\subsection{Searching for the Low-Abundance Elements}

The plasma state of the Fe knot constrained by our diagnostics predicts centroid energies of the Cr, 
Mn, and Ni \Ka\ emission to be 5480--5490\,eV, 5952--5960\,eV, and 7509--7514\,eV, respectively. 
Allowing the line centroid to vary within these ranges, we derive their flux upper limits 
relative to Fe \Ka\ as given in Table\,\ref{tab:fit-suzaku}. 
\cite{Miceli15} recently reported that the largest equivalent width (EW) of the Cr \Ka\ line
in this SNR is found at the Fe knot, based on their analysis of {\it XMM-Newton}/EPIC data. 
We constrain, however, the Cr/Fe EW ratio to $<$\,0.016, ruling out their detection ($0.033 \pm 0.015$). 
Although the HPD of the {\it Suzaku}/XRT is larger than that of {\it XMM-Newton} 
and thus some photons from the Fe knot are lost, this does not affect the line flux ratios 
because the XRT point spread function is almost independent of the X-ray energy \citep{Serlemitsos07}.
Moreover, even if some photons from the outer regions with a `normal' Cr/Fe ratio contribute to 
the Fe knot spectrum, this would rather increase the XIS-measured Cr/Fe ratio. 
We therefore consider that the lack of Cr emission in the Fe knot is robust.
\cite{Miceli15} also reported the detection of Ti and Mn \Ka\ lines from this region (though the latter is marginal). 
In Figure\,\ref{fig:spec-knot}(a), we show their best-fit model (normalized to the XIS spectrum) with the red line, 
where the EW (or flux) ratios among the lines they reported are assumed. Figure\,\ref{fig:spec-knot}(b) and 
\ref{fig:spec-knot}(c) show the residuals for our best-fit and Miceli et al.'s models, respectively. 
The latter clearly shows negative residuals at each line energy. 
We confirm that the addition of any lines significantly increases the $\chi ^2$ value 
(e.g., from 99 to 116 for Mn \Ka, with 106 degrees of freedom).

\begin{figure}[t]
  \begin{center}
	\includegraphics[width=8.2cm]{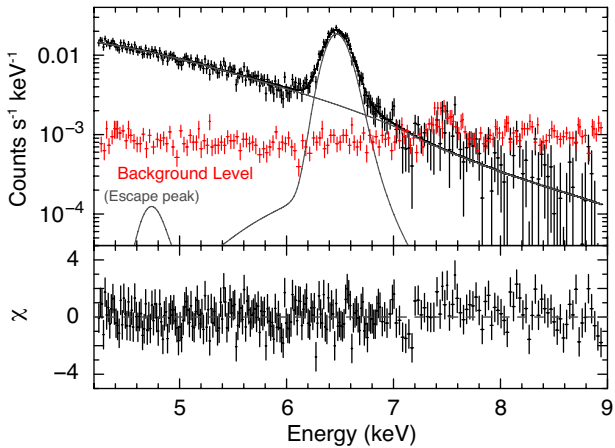}	
	\vspace{1mm}
\caption{{\it Chandra}/ACIS spectrum of the entire Fe knot in the 4.2--9.0\,keV band (black), fitted with 
a model consisting of a Gaussian (for the Fe \Ka\ emission) and a power low (for the continuum). 
The red data points are the background spectrum for the knot region, indicating that the source spectrum 
above 7\,keV is dominated by the background in the ACIS data. The structure around 4.8\,keV 
of the model (gray line) is a Si-escape peak of the Fe \Ka\ component (detector origin). 
The bottom panel shows the residual between the data and model.
  \label{fig:cxo-hard}}
  \end{center}
\end{figure}

\begin{figure*}[t]
  \begin{center}
  	\includegraphics[width=17.2cm]{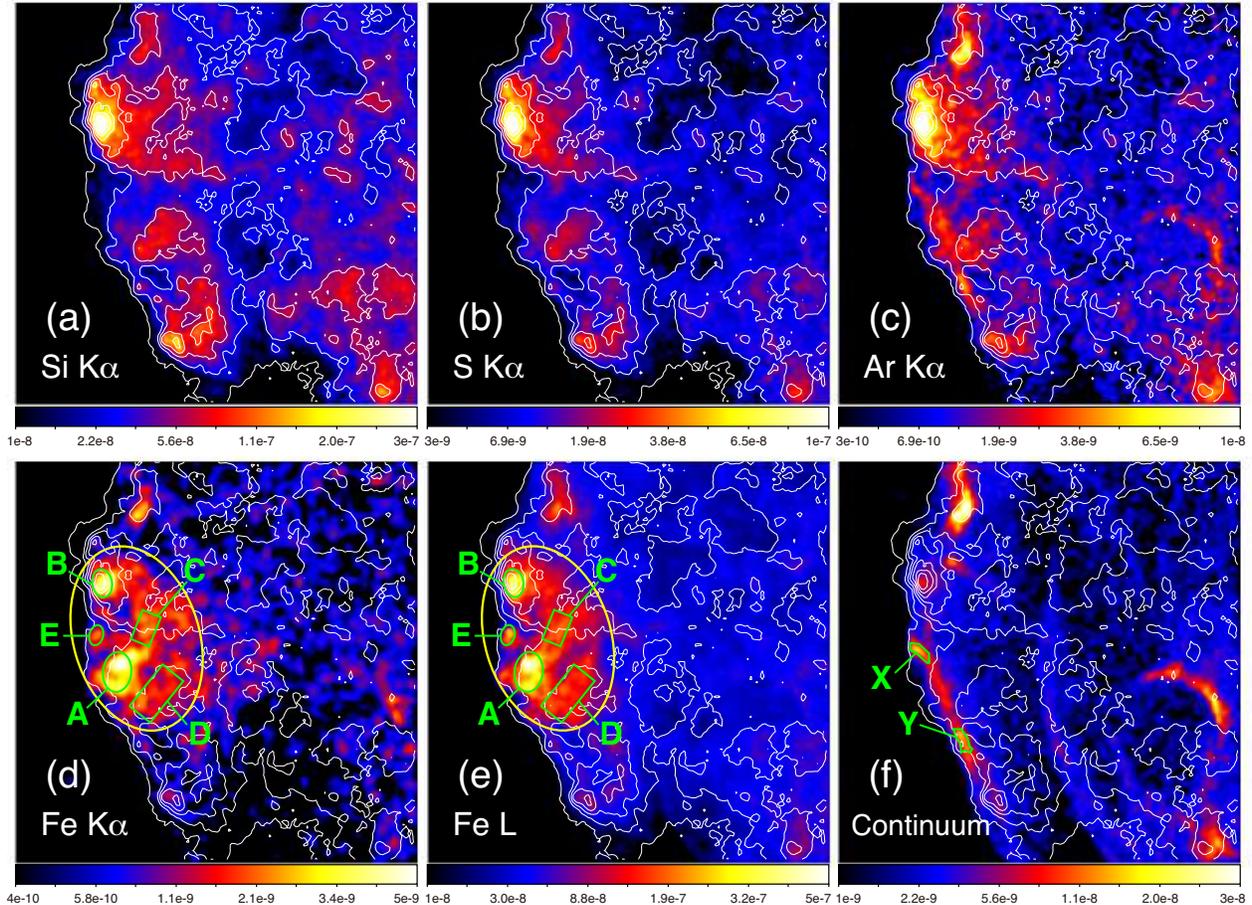}
	\vspace{1mm}
\caption{
 {\it Chandra}/ACIS images of the eastern region of {\it Tycho}'s SNR
 (the yellow square in Figure\,\ref{fig:image}) containing the Fe knot. 
 The energy band of each panel is following: (a) 1.8--1.92\,keV, (b) 2.4--2.52\,keV, (c) 3.07--3.18\,keV, 
 (d) 6.35-6.6\,keV, (e) 0.8--1.25\,keV, and (f) 4.2--6.0\,keV. 
 The unit of the color bar values is photons\,cm$^{-2}$\,s$^{-1}$. 
  The contours for the 1.80--1.92\,keV (Si \Ka) emission are overlaid on all the images. 
  The yellow ellipse encloses the entire Fe knot. 
  The spectra from the smaller regions (green) are analyzed in detail in \S2.6. 
  \label{fig:image-cxo}}
  \end{center}
\end{figure*}

We also analyze the {\it Chandra}/ACIS data to check the consistency. Figure\,\ref{fig:cxo-hard} 
shows an ACIS spectrum in the 4.2--9.0\,keV band extracted from the entire Fe knot  
(yellow ellipse in Figure\,\ref{fig:image-cxo}). A background spectrum is taken from 
a nearby region in the same CCD chip, which is also shown in Figure\,\ref{fig:cxo-hard}. 
Despite the relatively high background level (compared to the XIS), the obtained photon statistics 
are still high enough at the energies below 7\,keV, owing to the longer exposure and smaller 
photon-escape effect. We see, however, no clear feature of Cr or Mn \Ka\ emission. 
The upper limit of the Cr/Fe EW ratio is obtained to be 0.009 (and the flux ratio is $<$\,0.028), again, 
ruling out the {\it XMM-Newton} result. Note that the total effective exposure of the {\it XMM-Newton} 
data \citep{Miceli15} was only $\sim$\,125\,ks, and they analyzed only EPIC-pn spectra, 
of which instrumental background contains Cr fluorescence lines and is largely position 
dependent\footnote{https://heasarc.gsfc.nasa.gov/docs/xmm/uhb/epicintbkgd.html}. 
Incomplete background subtraction may explain their detection of a Cr line. 
Our analysis is based on the XIS-FI data with the lowest, stable background level plus  
the best available energy resolution, together with the ACIS deep observations.  
The result is therefore considered to be more reliable than any previous measurements.

The detection of a Ti line is also questionable from a theoretical consideration. 
While the reported centroid of $\sim$\,4.93\,keV corresponds to the Ly$\alpha$ line of {\it H-like} Ti, 
based on the plasma conditions in the Fe knot ($n_et \approx 1.2 \times 10^{10}$\,cm$^{-3}$\,s) 
the Ti \Ka\ centroid is expected to be $\sim$\,4.59\,keV (indicated by the red triangle in Figure\,\ref{fig:spec-knot}(c)).
Since the majority of the ejecta in {\it Tycho}'s SNR is in the He-like or even lower charge states 
\citep[e.g.,][]{Badenes06,Yamaguchi14b}, it is unlikely that only Ti is in the highly-ionized H-like state. 
Further studies might be required to conclude how this possibly artificial 4.9-keV line is formed.


\subsection{Chandra High-Resolution Images}

In Figure\,\ref{fig:image-cxo}, we show narrow band ACIS images of the region around 
the Fe knot at the energies of (a) Si \Ka, (b) S \Ka, (c) Ar \Ka, (d) Fe \Ka, 
(e) Fe {\footnotesize XVII}--{\footnotesize XIX} L-shell blend, and (f)  
4.2--6.0 keV continuum emission. Each image is overplotted with the contours of the Si \Ka\ emission.
The surface brightness distribution of the Si, S, and Ar \Ka\ emission (i.e., the IMEs) 
is quite similar to one another. On the other hand, the Fe emission 
(both \Ka\ and L-shell blend) exhibits a distinctly different morphology from that of the IMEs. 
A thin filamentary feature found in the continuum band is similar to the nonthermal rims seen elsewhere 
in this remnant and suggests that we are seeing the forward shock projected on the Fe knot emission.

\begin{table}[t]
\begin{center}
\caption{Fe \Ka\ flux and size of each subregion.
  \label{tab:subreg}}
    \begin{tabular}{lcccc}
\hline \hline
Knot & Fe \Ka\ flux & \multicolumn{2}{c}{Angular size}  & Volume \\
\cline{3-4}
~ & [$10^{-6}$\,ph\,cm$^{-2}$\,s$^{-1}$] & [arcsec$^2$] & [pc$^2$] & [$10^{54}$\,cm$^3$] \\
\hline
A & $4.58 \pm 0.41$ & 231 & 0.0486 & 1.00 \\
B & $1.94 \pm 0.28$ & 113 & 0.0239 & 0.494 \\
C & $1.79 \pm 0.25$ & 160 & 0.0338 & 0.698 \\
D & $3.03 \pm 0.33$ & 364 & 0.0770 & 1.59 \\
E & $0.54 \pm 0.19$ & 57.1 & 0.0121 & 0.250 \\
\hline
Total$^{\ast}$ & $35.2 \pm 1.2$ & 4820 & 1.01 & 21.0 \\
\hline
\end{tabular}
\tablecomments{
The distance to the SNR is assumed to be 3\,kpc. \\
$^{\ast}$``Total'' is not the sum of Knots A--E, but gives the flux from 
the whole Fe knot and the size of the yellow ellipse in Figure\,\ref{fig:image-cxo}.
}
\end{center}
\end{table}

Based on its morphology, the Fe knot can be divided into substructures. 
The brightest spot, indicated as ``A'' in Figure\,\ref{fig:image-cxo}(d) 
or \ref{fig:image-cxo}(e), is found near the middle of the Fe knot. 
This feature seems physically distinct from the surrounding IME blobs. 
The second brightest peak ``B'', on the other hand, spatially coincides with the IME emission.  
The other subregions, ``C'', ``D'', and ``E'', are also bright in the Fe bands 
but not associated with the morphology of the IMEs. In short, only Knot~B may have a different 
chemical composition from the others. The Fe \Ka\ line flux from and physical size of each subregion 
are listed in Table\,\ref{tab:subreg}. The emitting volume (also given in Table\,\ref{tab:subreg}) 
is roughly estimated assuming a plasma depth of 0.7\,pc, which corresponds to an angular size of $48''$ 
at the nominal distance of 3\,kpc. Readers should be warned that this depth and thus the estimated 
volume have relatively large uncertainty. The average surface brightness of 
the Fe \Ka\ emission in the whole Fe knot (yellow ellipse in Figure\,\ref{fig:image-cxo}) 
is $7.3 \times 10^{-9}$\,photons\,cm$^{-2}$\,s$^{-1}$\,arcsec$^{-2}$, 
about 37\% of that at the brightest Knot~A.

\subsection{Small-Scale Spectra}

\begin{figure}[t]
  \begin{center}
	\includegraphics[width=8.2cm]{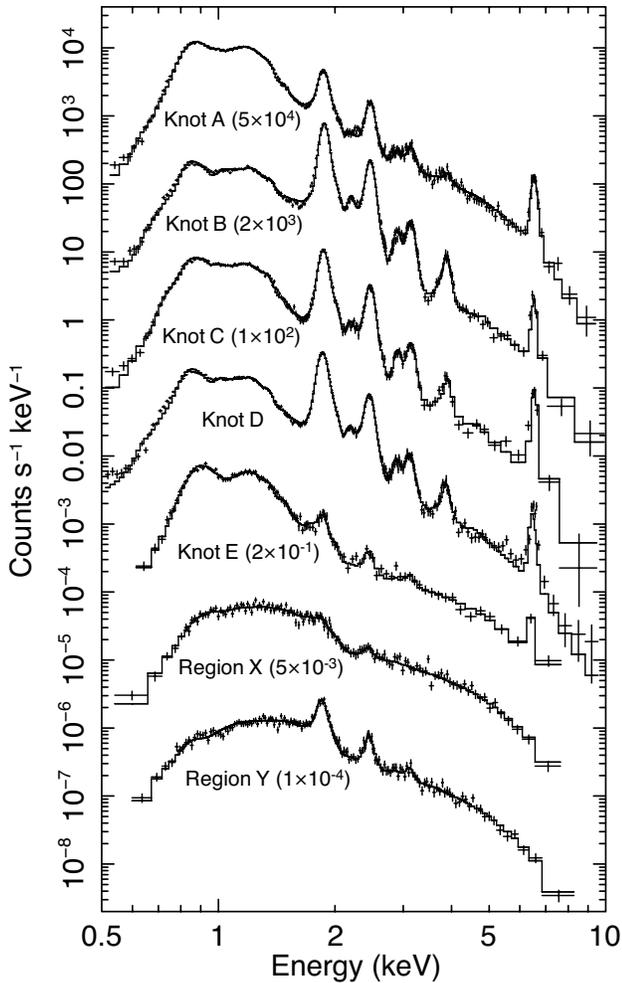}	
	\vspace{1mm}
\caption{ACIS spectra (background subtracted) taken from the small regions given 
in Figure\,\ref{fig:image-cxo}. The normalization of each spectrm is adjusted by multiplying 
the factor given in the panel. 
\label{fig:spec-allcxo}}
  \end{center}
\end{figure}

An ACIS spectrum from each subregion is extracted and shown in Figure\,\ref{fig:spec-allcxo}, 
where a background taken from the same CCD chip is subtracted. We also extract nonthermal 
spectra from the regions ``X'' and ``Y'' in Figure\,\ref{fig:image-cxo}(f) to investigate 
the characteristic continuum shape (i.e., photon index) of spectra around the knot. 
We fit these nonthermal spectra with an absorbed power law plus a pure-metal NEI model 
(see below for details) with free abundances of Si, S, Ar, and Fe that appear in the spectra. 
We obtain \NH\ = $7.9_{-0.7}^{+0.9} \times 10^{21}$\,cm$^{-2}$ and $\Gamma$ = 2.58 $\pm$ 0.06 
for Region X, and \NH\ = $9.4_{-1.2}^{+0.9} \times 10^{21}$\,cm$^{-2}$ and $\Gamma$ = 
$2.68_{-0.08}^{+0.07}$ for Region Y.  Both parameters overlap between the two regions, 
and are consistent with the typical values in this SNR \citep{Cassam07}.

\begin{figure}[t]
  \begin{center}
	\includegraphics[width=8.2cm]{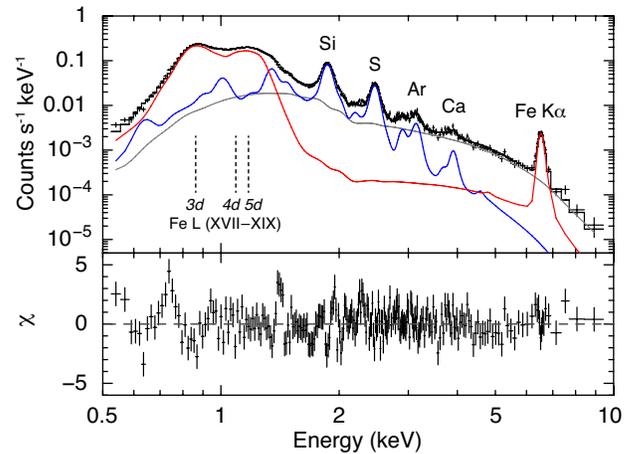}	
	\vspace{1mm}
\caption{The ACIS spectrum of Knot A. The best-fit models for the Fe knot, IME, and 
power-law components are given in the red, blue, and gray lines, respectively. 
The bottom panel shows the residual between 
the data and model.
  \label{fig:spec-cxo}}
  \end{center}
\end{figure}

Figure\,\ref{fig:spec-cxo} shows the background-subtracted spectrum of Knot A, the same as shown
in Figure\,\ref{fig:spec-allcxo}. No feature of Cr, Mn, or Ni emission is confirmed in this spectrum.
Owing to the relatively small amount of contamination from the IME component in this subregion 
(see Figure\,\ref{fig:image-cxo}), we see prominent features of the L-shell blend 
of Fe\,{\footnotesize XVII}, Fe\,{\footnotesize XVIII}, and Fe\,{\footnotesize XIX} at $\sim$\,0.83\,keV 
($3d$$\rightarrow$$2p$) and $\sim$\,1.1\,keV (transitions from $n \geq 4$ to $n = 2$). 
This indicates that the L-shell emission from the Fe knot is dominated by Ne-like and F-like Fe, 
consistent with the origin of the Fe \Ka\ fluorescence (\S2.3). 
The identical origin of the Fe L and \Ka\ emission is also implied from their similar morphology as 
seen in Figure\,\ref{fig:image-cxo}(d) and \ref{fig:image-cxo}(e). This result allows us to determine 
the characteristic electron temperature of the Fe knot using the L-shell/K-shell flux ratio, 
because their line emissivities strongly depend on the internal energy (i.e., temperature) 
of free electrons.

\begin{table}[t]
\begin{center}
\caption{The best-fit spectral parameters for Knot A.
  \label{tab:fit-cxo}}
  \begin{tabular}{llc}
\hline \hline
Components & Parameters & Values \\
\hline
Absorption & \NH\ [$10^{21}$\,cm$^{-2}$] & $8.1 \pm 0.1$ \\
Fe knot & $kT_e$ [keV] & $8.1_{-0.4}^{+0.6}$ \\
(low-$n_et$)   & $n_et$ [$10^{10}$\,cm$^{-3}$\,s] & $1.26 \pm 0.02$ \\
~              & $n_e n_{\rm Fe} V$ [$10^{51}$\,cm$^{-3}$] & $3.93 \pm 0.03$ \\
IME          & $kT_e$ [keV] & $1.5_{-0.1}^{+0.3}$  \\
(high-$n_et$)   & $n_et$ [$10^{10}$\,cm$^{-3}$\,s] & $5.42_{-0.55}^{+0.21}$ \\ 
~              & $n_e n_{\rm Si} V$ [$10^{51}$\,cm$^{-3}$] &  $3.49_{-0.04}^{+0.05}$  \\
~  & $n_{\rm Ne}/n_{\rm Si}$ & $0.98 \pm 0.09$  \\
~  & $n_{\rm Mg}/n_{\rm Si}$ & $0.49 \pm 0.02$  \\
~  & $n_{\rm S}/n_{\rm Si}$ & $0.65 \pm 0.03$  \\
~  & $n_{\rm Ar}/n_{\rm Si}$ & $0.13 \pm 0.02$  \\
~  & $n_{\rm Ca}/n_{\rm Si}$ & $0.11 \pm 0.04$  \\
Power law & $\Gamma$ & $2.62_{-0.04}^{+0.03}$ \\
~ & Normalization$^{a}$ & $3.20_{-0.70}^{+0.83}$ \\
\hline
$\chi ^2 /d.o.f$ & & 380/228 \\
\hline
\end{tabular}
\tablecomments{
$^{a}$Differential photon flux at 1\,keV in the unit of $10^{-5}$~keV$^{-1}$~cm$^{-2}$~s$^{-1}$.
}
\end{center}
\end{table}

\begin{table*}[t]
\begin{center}
\caption{The best-fit parameters for the Chandra spectra of the Fe knot subregions.
  \label{tab:fit-cxoall}}
  \begin{tiny}
    \begin{tabular}{lcccccccccc}
\hline \hline
\multicolumn{9}{c}{I. Nonthermal continuum model} \\
\hline
Knot & \NH\ & $kT_e$\,(Fe) & $n_e t$\,(Fe) & $n_e n_{\rm Fe} V$ & 
$kT_e$\,(IME) & $n_e t$\,(IME) & $n_e n_{\rm Si} V$ & $\Gamma$ & $\chi^2$/d.o.f. & $\chi^2$/d.o.f. \\
~ & [$10^{21}$\,cm$^{-2}$] & [keV] & [$10^{10}$\,cm$^{-3}$\,s] &  [$10^{51}$\,cm$^{-3}$] & 
[keV] & [$10^{10}$\,cm$^{-3}$\,s] &  [$10^{51}$\,cm$^{-3}$] & & (Full band)$^{a}$ & (Fe L/K)$^{b}$  \\
\hline
A & $8.1 \pm 0.1$ & $8.1_{-0.4}^{+0.6}$ & $1.26 \pm 0.02$ & $3.93 \pm 0.03$ &
$1.5_{-0.1}^{+0.3}$ & $5.42_{-0.55}^{+0.21}$ & $3.49_{-0.04}^{+0.05}$ & $2.62_{-0.04}^{+0.03}$ 
& 380/228 & 76/67 \\
B & $6.8 \pm 0.1$ & $8.9_{-0.4}^{+0.2}$ & $1.01_{-0.02}^{+0.03}$ & $1.35_{-0.03}^{+0.01}$ &
$14 \pm 0.8$ & $1.74 \pm 0.02$ & $8.32_{-0.04}^{+0.08}$ & $2.85_{-0.05}^{+0.07}$ 
& 1080/207 & 68/47 \\
C & $8.5 \pm 0.1$ & $8.6_{-0.8}^{+0.3}$ & $1.33 \pm 0.02$ & $1.57_{-0.02}^{+0.03}$ &
$1.5 \pm 0.1$ & $5.21_{-0.16}^{+0.30}$ & $5.02_{-0.07}^{+0.05}$ & $2.62_{-0.08}^{+0.12}$ 
& 336/163 & 48/43 \\
D & $6.5 \pm 0.1$ & $8.6_{-0.3}^{+0.2}$ & $1.06 \pm 0.02$ & $2.06 \pm 0.02$ &
$1.4 \pm 0.1$ & $5.03_{-0.15}^{+0.11}$ & $13.9 \pm 0.1$ & $2.62 \pm 0.04$ 
& 812/224 & 125/58 \\
E & $8.5 \pm 0.7$ & $6.0_{-1.4}^{+2.0}$ & $1.59 \pm 0.06$ & $0.61_{-0.08}^{+0.10}$ &
$3.4_{-1.8}^{+9.5}$ & $3.20_{-0.77}^{+3.15}$ & $0.14_{-0.05}^{+0.03}$ & $2.49 \pm 0.18$ 
& 129/103 & 43/40 \\
\hline
\multicolumn{9}{c}{II. Thermal continuum model} \\
\hline
Knot & \NH\ & $kT_e$\,(Fe) & $n_e t$\,(Fe) & $n_e n_{\rm Fe} V$ & 
$kT_e$\,(IME) & $n_e t$\,(IME)& $n_e n_{\rm H} V$ & $n_{\rm Si}/n_{\rm H}$ & $\chi^2$/d.o.f. & $\chi^2$/d.o.f. \\
~ & [$10^{21}$\,cm$^{-2}$] & [keV] & [$10^{10}$\,cm$^{-3}$\,s] &  [$10^{51}$\,cm$^{-3}$] & 
[keV] & [$10^{10}$\,cm$^{-3}$\,s] &  [$10^{55}$\,cm$^{-3}$] & [$\times$\,$10^{-5}$]  & 
(Full band)$^{a}$ & (Fe L/K)$^{b}$  \\
\hline
A & $7.9 \pm 0.1$ & $8.7_{-0.4}^{+0.2}$ & $1.18 \pm 0.02$ & $3.42 \pm 0.03$ &
$3.0 \pm 0.1$ & $2.86_{-0.10}^{+0.06}$ & $3.46 \pm 0.03$ & $7.1 \pm 0.2$ 
& 379/228 & 80/68 \\
B & $6.7 \pm 0.1$ & $8.7_{-0.5}^{+0.3}$ & $0.91 \pm 0.02$ & $1.09 \pm 0.02$ &
$4.0 \pm 0.1$ & $1.95_{-0.04}^{+0.02}$ & $0.93_{-0.07}^{+0.01}$ & $102 \pm 1$ 
& 960/207 & 68/48 \\
C & $8.8 \pm 0.1$ & $8.9_{-0.5}^{+0.8}$ & $1.19_{-0.06}^{+0.03}$ & $1.38_{-0.21}^{+0.04}$ &
$2.2_{-0.1}^{+0.2}$ & $3.10_{-0.27}^{+0.11}$ & $0.69_{-0.16}^{+0.08}$ & $57.7 \pm 3.7$ 
& 328/163 & 50/44 \\
D &  $7.5 \pm 0.1$ & $8.6_{-0.6}^{+0.4}$ & $1.26 \pm 0.02$ & $1.76 \pm 0.03$ &
$2.4 \pm 0.1$ & $2.53_{-0.04}^{+0.05}$ & $2.24 \pm 0.02$ & $47.1_{-0.4}^{+0.7}$ 
& 1350/224 & 101/59 \\
E & $8.0 \pm 0.8$ & $6.5_{-1.8}^{+2.4}$ & $1.45_{-0.96}^{+0.10}$ & $0.48_{-0.11}^{+0.13}$ &
$3.6_{-0.7}^{+0.9}$ & $3.13_{-0.39}^{+0.55}$ & $0.68_{-0.11}^{+0.15}$ & $2.2 \pm 0.6$ 
& 128/103 & 45/41 \\
\hline
\end{tabular}
\tablecomments{
$^{a}$The $\chi^2$/d.o.f.\ values for the full-band fitting. \ 
$^{b}$The $\chi^2$/d.o.f.\ values for the Fe L- and K-shell bands only (see text).  
}
\end{tiny}
\end{center}
\end{table*}

We fit the Knot A spectrum with an absorbed plasma model consisting of two NEI components, 
one for the Fe emission and the other for the IMEs, for which different origins are suggested 
from their morphology (\S2.5). 
We also add a power-law component to reproduce the continuum that is likely dominated by 
nonthermal emission. The free parameters are listed in Table\,\ref{tab:fit-cxo}. 
For the thermal NEI components, we assume pure-metal plasmas without any admixture of 
other elements including hydrogen \citep[see][for technical details]{Yamaguchi08b}.  
The emission measure of the Fe knot component is, therefore, defined as a product of the electron 
and Fe densities and the emitting volume, $n_e n_{\rm Fe} V$, instead of ordinary $n_e n_{\rm H} V$ 
(and similarly $n_e n_{\rm Si} V$ is defined for the IME component). 
We compensate for `missing lines' in the current {\it AtomDB}, such as high-level ($n > 5$) L-shell 
transitions of Fe\,{\footnotesize XVIII}, by adding line components around  $\sim$\,1.2\,keV. 
The best fit is then obtained with the parameters given in Table\,\ref{tab:fit-cxo}. 
The photon index of the nonthermal component is consistent with those in Regions X and Y. 
As expected from the presence of Fe\,{\footnotesize XVII--XIX} L-shell emission, 
the ionization timescale obtained for the Fe knot component is consistent with the value determined 
from the Fe K diagnostics in \S2.3. The distinct plasma parameters (i.e., $kT_e$ and $n_et$) 
between the Fe and IME components confirm their independent origins. 
For the rest of this section, we take the position that the IME emission in the spectrum of 
Knot A is due to projection and that the Knot is of pure Fe composition.

To assess the uncertainty in the electron temperature of the Fe knot, 
we refit the spectrum assuming a thermal origin for the continuum emission; 
we replace the pure-metal IME component with a hydrogen-dominant NEI plasma, and remove 
the power law from the model. This model also reproduces the hard X-ray continuum well. 
The best-fit parameters are given in the lower segment of Table\,\ref{tab:fit-cxoall}; the fit shows 
that the $kT_e$ value of the Fe component is insensitive to other model parameters. 
We also analyze the spectra from the other subregions using the models described above. 
The results from both (I) the nonthermal continuum model and (II) the thermal continuum model 
are summarized in Table\,\ref{tab:fit-cxoall} (upper and lower segments, respectively). 
The obtained $\chi^2$/d.o.f.\ values are large particularly for the spectra with a high count rate 
(e.g., Knot B). In those, the discrepancy between the data and model is mainly found around 
the IME emission, possibly due to intrinsic line broadening and/or incomplete calibration. 
We, however, do not try to find a better fit by introducing more complicated models, because 
our purpose here is exclusively to determine the electron temperature of the Fe knot component. 
After the best fit is obtained with the full-band spectra, we re-fit only the Fe L (0.7--1.3\,keV) and 
Fe K (above 4.2\,keV) bands by fixing the parameters of the IME components. The results are then 
obtained with reasonable $\chi^2$/d.o.f.\ values (the rightmost column of Table\,\ref{tab:fit-cxoall}). 
The electron temperature does not change significantly from the original values given in the table.

We find no significant spatial variation in the electron temperature of the Fe knot, of which typical 
values are consistent with independent analysis results from \cite{Sato16}. 
Although the temperature is marginally lower at Knot E, its Fe \Ka\ flux is only 1.5\% of the total flux 
from the knot (Table\,\ref{tab:subreg}). Therefore, we consider $kT_e$ = 7--10\,keV  
to be a good approximation for the electron temperature of the Fe knot component. 
Distinct plasma parameters between the Fe and IMEs are also confirmed in each subregion,  
with the possible exception of Knot B, where the $n_et$ values for the two components 
are relatively close to each other. This similarity is consistent with the indication from the imaging study; 
Fe might be mixed with the IMEs to a certain degree in this subregion. 
{The Knot D spectrum shows an emission-like feature around $\sim$\,5.5\,keV. 
We fit this feature with an additional Gaussian to assess the possibility of Cr emission. 
We find that its centroid energy (5550 $\pm$ 40\,eV) is significantly higher than 
the expected value (5480--5490\,eV) for the plasma condition of the Fe knot (see \S 2.4), 
suggesting that this feature, if real, is associated with the higher-$n_et$ IME component. 
This interpretation is supported by the fact that the Knot D region spatially coincides with 
a peak of the IME emission (see Figure\,\ref{fig:image-cxo}). 
A further investigation is left for future work.

\subsection{Fe Mass and Mass Ratios}

At the electron temperature and ionization timescale constrained above, 
the Fe \Ka\ line flux from the entire Fe knot ($3.5 \times 10^{-5}$\,photons\,cm$^{-2}$\,s$^{-1}$) 
corresponds to an emission measure $n_e n_{\rm Fe} V$ of $3.0 \times 10^{52}$\,cm$^{-3}$. 
The pure-Fe composition leads to a relationship between the electron and Fe densities 
of $n_e$ = $\left< z_{\rm Fe} \right> n_{\rm Fe}$ $\approx$ 17\,$n_{\rm Fe}$. 
Using this relationship and the emitting volume given in Table\,\ref{tab:subreg}, we obtain 
an average ion density of $n_{\rm Fe}$ $\approx$ $7.5 \times 10^{-3}$\,cm$^{-3}$. 
The total Fe mass is roughly estimated as $M_{\rm Fe}$ $\approx$ $m_{\rm Fe} n_{\rm Fe} V$ 
= $7.4 \times 10^{-3}$\Msun, where $m_{\rm Fe}$ is the mass of a single Fe nucleus. 
We note, however, that there is a substantial uncertainty in this estimate, because the actual 
density distribution in the knot is highly heterogeneous (see Figure\,\ref{fig:image-cxo}). 
Moreover, both $n_{\rm Fe}$ and $M_{\rm Fe}$ could be significantly lower if electrons ionized 
from other elements contribute to the Fe emission.

Finally, we refit the {\it Suzaku}/XIS spectrum of the Fe knot (Figure\,\ref{fig:spec-knot}), 
with a realistic NEI model by constraining the electron temperature to that determined 
through the spectral analysis of the {\it Chandra} data (i.e., $kT_e$ = 7--10\,keV).  
Such a fit yields upper limits for the mass ratios of $M_{\rm Cr}/M_{\rm Fe}$ $<$ 0.023, 
$M_{\rm Mn}/M_{\rm Fe}$ $<$ 0.012, and $M_{\rm Ni}/M_{\rm Fe}$ $<$ 0.029 
at the 90\% confidence level with a good $\chi^2$/d.o.f.\ value of 97/103. 
In this step, we apply a single pure-metal plasma model based on the latest {\it AtomDB} 
(which contains all the elements from H to Zn) to the emission lines and a power law to the continuum. 
The temperature dependence of the mass ratios is relatively small; 
the upper limit values change only $\lesssim$\,10\% within the constrained electron temperature. 
The same analysis is performed for the ACIS spectrum of Figure\,\ref{fig:cxo-hard}, but only below 7.0\,keV 
is used since the ACIS data above that energy is dominated by the background. 
We obtain $M_{\rm Cr}/M_{\rm Fe}$ $<$ 0.021 and $M_{\rm Mn}/M_{\rm Fe}$ $<$ 0.010 
with $\chi^2$/d.o.f.\ = 154/184, consistent with the XIS results. 
A simultaneous fit of both XIS and ACIS spectra does not change the mass ratios significantly from 
the XIS-only results, probably because the signal-to-noise ratio is much higher in the XIS spectrum. 
We emphasize that this mass ratio estimate is independent of the density distribution in 
the Fe knot and the distance to the SNR. Therefore, these uncertainties do not affect
our main conclusions.

\section{Interpretation and Discussion}

Based on the combined analysis of the {\it Suzaku} and {\it Chandra} deep observations 
of {\it Tycho}'s SNR, we have inferred the physical properties of the Fe knot. 
Its Fe \Ka\ and L-shell emission is reasonably well represented by a single `pure-Fe' NEI 
plasma component. The relationship between its ionization timescale and the physical location 
(distance from the SNR center) of this knot differs from the trend found elsewhere in  
the SNR shell (Figure\,\ref{fig:r-nt}). These findings strongly suggest that the Fe knot is 
an ejecta clump physically isolated from the bulk of the reverse-shocked material. 
Emission from the secondary Fe-peak elements (i.e., Cr, Mn, and Ni) is not detected even in 
the sensitive {\it Suzaku} data, in strong contrast to the other Fe-rich regions. 
The {\it Chandra} high-resolution images and spectra indicate that the Fe and IMEs have 
different spatial distributions. We thus conclude that the Fe knot mostly consists of ``pure Fe'', 
with no mixture of any other metals.

Previous studies of {\it Tycho}'s SNR agree on its origin as a typical SN Ia from 
a Chandrasekhar-mass ($M_{\rm Ch}$) C+O white dwarf \citep{Badenes06,Krause08}. 
In such an explosion, Fe and its parent nuclei, like $^{56}$Ni, are synthesized in the nuclear 
burning regime of either incomplete Si burning, $\alpha$-rich freeze-out NSE ($\alpha$-NSE), 
normal NSE, or neutron-rich NSE (n-NSE), depending on the peak temperature ($T_{\rm peak}$) 
and density ($\rho_{\rm peak}$) of the burning materials \citep[e.g.,][]{Thielemann86,Iwamoto99}. 
The secondary Fe-peak elements are co-synthesized with Fe in all of these regimes. 
Therefore, their absence from the Fe knot is surprising. Incomplete Si burning 
takes place at $4.5\times 10^9 \lesssim T_{\rm peak}\,[{\rm K}] \lesssim 5.5 \times 10^9$, 
and yields Cr and Mn together with Fe. 
At $T_{\rm peak} \gtrsim 5.5 \times 10^9$\,K, $\alpha$-NSE burning becomes dominant 
and mainly produces Fe and stable Ni.  
NSE burning (normal freeze-out) occurs in the higher-density/lower-entropy regions with 
$\rho_{\rm peak} \gtrsim 2 \times 10^8$\,g\,cm$^{-3}$ and $T_{\rm peak} \gtrsim 5.5 \times 10^9$\,K, 
leaving a larger amount of $^{55}$Co which decays into Mn \citep[e.g.,][]{Seitenzahl13b}.
In an even denser environment, or the innermost region of a $M_{\rm Ch}$ SN Ia, 
efficient electron capture takes place, so a large number of neutron-rich nuclei are directly
produced (n-NSE). For this reason, a high Ni/Fe mass ratio, $M_{\rm Ni}/M_{\rm Fe} \gtrsim 0.1$, 
is expected for this burning regime \citep[e.g.,][]{Yamaguchi15}; this expectation conflicts with 
our inferred upper limit of $M_{\rm Ni}/M_{\rm Fe}$ $<$ 0.029.

\subsection{Constraining the Burning Regime}

\begin{figure}[p]
  \begin{center}
	\includegraphics[width=8.3cm]{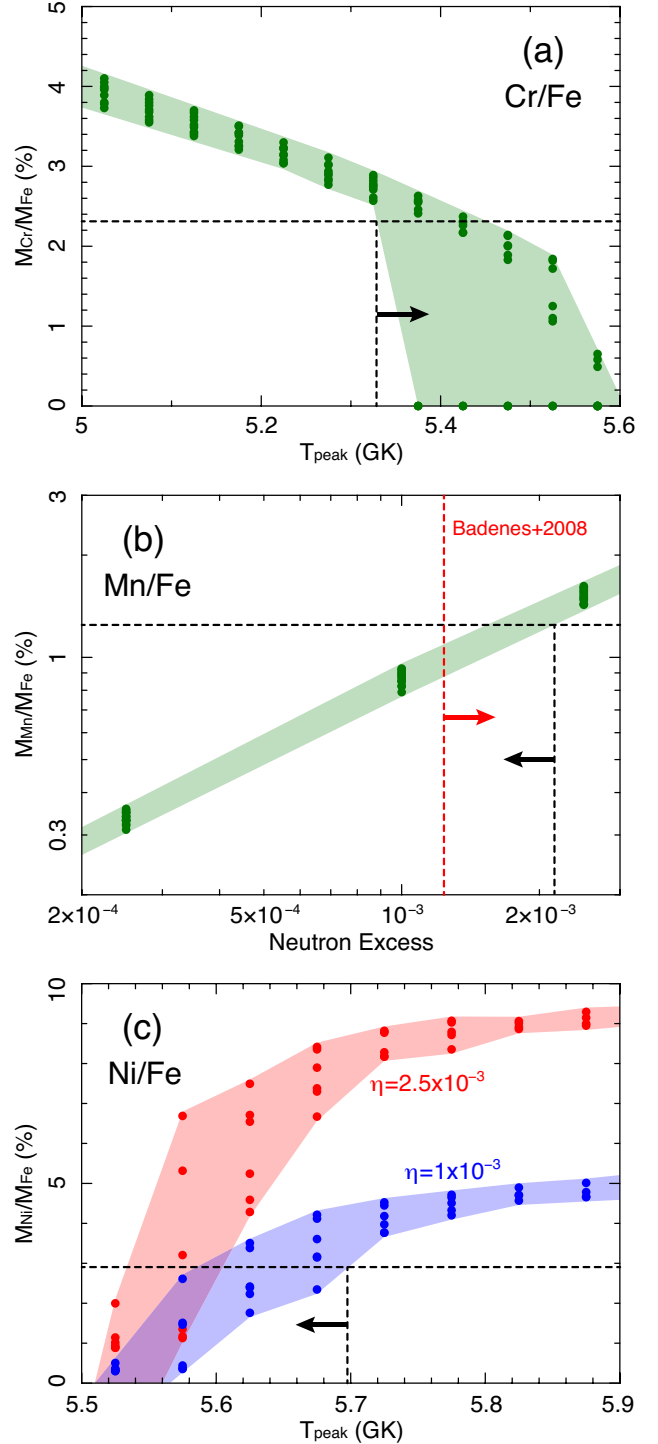}	
	\vspace{1mm}
\caption{
(a) Relationship between $M_{\rm Cr}/M_{\rm Fe}$ and $T_{\rm peak}$ predicted by 
the delayed-detonation $M_{\rm Ch}$ SN Ia models presented in \cite{Yamaguchi15}. 
The horizontal and vertical dashed lines indicate the upper limit for the mass ratio 
observed in the Fe knot 
and the corresponding conservative lower limit for $T_{\rm peak}$, respectively. 
(b) Relationship between $M_{\rm Mn}/M_{\rm Fe}$ and $\eta$ expected for the burning cells 
with $5.3\times 10^9 \lesssim T_{\rm peak}\,[{\rm K}] \lesssim 5.6 \times 10^9$ of the same models 
as panel (a). The red vertical dashed line indicates the lower limit for $\eta$ determined by 
\cite{Badenes08b} for this SNR. 
(c) Relationship between $M_{\rm Ni}/M_{\rm Fe}$ and $T_{\rm peak}$ from the same models 
with different $\eta$ values of $1 \times 10^{-3}$ (blue) and $2.5 \times 10^{-3}$ (red). 
The observed mass ratio gives the upper limit for $T_{\rm peak}$ indicated by the vertical dashed line. 
   \label{fig:massratio}}
  \end{center}
\end{figure}

In order to constrain the burning regime that might have produced the Fe knot, 
we investigate grids of the standard delayed-detonation SN Ia models presented in 
\cite{Bravo12} and \cite{Yamaguchi15}. The progenitor is assumed to be a $M_{\rm Ch}$ 
C+O white dwarf (suggested by previous work) with identical mass fractions of C and O. 
Figure\,\ref{fig:massratio}(a) shows the predicted Cr/Fe mass ratio as a function of $T_{\rm peak}$, 
derived from relevant grids of these models. Although the values are calculated using 
a one-dimensional code \citep{Bravo12}, the ranges of 
$\rho_{\rm peak}$ and neutron excess ($\eta$ $\equiv$ $1 - 2 \left< Z_A / A \right>$, 
where $Z_A$ and $A$ are the atomic number and mass number, respectively) are reasonably covered. 
The excess neutrons are assumed to all come from $^{22}$Ne \citep[i.e., the metallicity effect:][]{Timmes03}, 
and hence the neutronization due to the pre-explosion carbon simmering is neglected \citep{Piro08,Martinez16}. 
We find that the Cr yield is almost independent of $\eta$, and the Cr/Fe ratio monotonically decreases 
as $T_{\rm peak}$ increases. The observed upper limit is consistent with the models 
only when $T_{\rm peak} \gtrsim 5.3 \times 10^9$\,K.

The Mn yield from incomplete Si burning strongly depends on $\eta$, in addition to $T_{\rm peak}$
\citep{Badenes08b}. We thus investigate in Figure\,\ref{fig:massratio}(b) the relationship between 
$M_{\rm Mn}/M_{\rm Fe}$ and $\eta$, predicted for the $T_{\rm peak}$ range constrained by 
the inferred Cr/Fe mass ratio. 
We find that the inferred $M_{\rm Mn}/M_{\rm Fe}$ upper limit requires $\eta \lesssim 2 \times 10^{-3}$. 
On the other hand, \cite{Badenes08b} determined its lower limit to be $\sim$\,$1.2 \times 10^{-3}$, 
using the Mn/Cr mass ratio derived from the integrated ejecta spectrum. 
Our result does not conflict with theirs. 
If the effect of carbon simmering is negligible, then the relationship between $\eta$ and 
the progenitor's metallicity $Z$ is given as $\eta$ = $0.101 \times Z$ \citep{Timmes03}. 
Therefore, the combined constraint from \cite{Badenes08b} and this work can be converted to 
$Z/Z_{\odot}$ = 0.9--1.5, using the up-to-date solar metallicity of $Z_{\odot}$ = 0.014 \citep{Asplund09}. 
This combination should be valid as long as the progenitor purely consists of C, O, and $^{22}$Ne, 
since the elemental composition (or $\eta$) in a $M_{\rm Ch}$ C+O white dwarf
is naturally expected to be uniform \citep[e.g.,][]{Iwamoto99}.

Finally, in Figure\,\ref{fig:massratio}(c) we plot the relationship among $M_{\rm Ni}/M_{\rm Fe}$, 
$T_{\rm peak}$, and $\eta$, in the $\alpha$-NSE burning regime using the same nucleosynthesis models. 
For the low $\eta$ case (blue), the predicted $M_{\rm Ni}/M_{\rm Fe}$ ratio range matches the observed 
upper limit at $T_{\rm peak} \lesssim 5.7 \times 10^9$\,K. The high $\eta$ case (red) requires an even lower 
temperature of $< 5.6 \times 10^9$\,K. In this temperature range, however, the predicted Mn/Fe ratio 
exceeds the inferred upper limit, unless $\eta \lesssim 2 \times 10^{-3}$ (Figure\,\ref{fig:massratio}(b)). 
Therefore, the high $\eta$ case is rejected also from the Ni/Fe mass ratio.

To summarize, {\it Tycho}'s Fe knot must have originated from 
either Si burning or $\alpha$-NSE regimes, probably close to their boundary with $T_{\rm peak}$ 
$\approx$ (5.3--5.7) $\times 10^9$\,K, if the $M_{\rm Ch}$ C+O progenitor scenario is the case. 
The neutron excess in the pre-explosion white dwarf is accordingly constrained to be 
(1.2--2.0)\,$\times$\,$10^{-3}$, which corresponds to the metallicity very close to the solar value. 
The normal freeze-out and n-NSE regimes are completely ruled out, indicating that 
the Fe knot does not come from near the core of a $M_{\rm Ch}$ white dwarf.

\subsection{Comparison with a Multi-Dimensional Model}

\begin{figure}[t]
  \begin{center}
	\includegraphics[width=8.7cm]{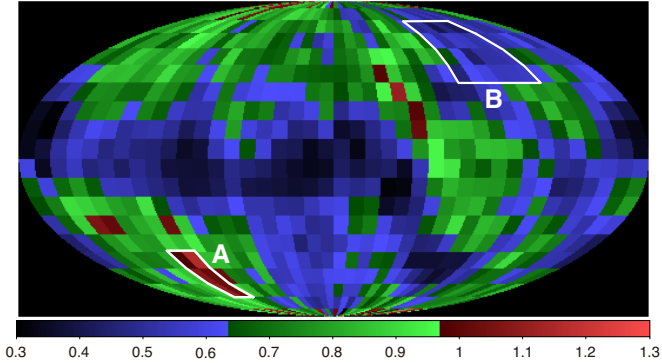}	
	\vspace{1mm}
\caption{
$\left<R_{\rm Fe}\right>$/$\left<R_{\rm Si}\right>$ for each small cell in 
the spherical coordinate system predicted by the N100 model of \cite{Seitenzahl13a}. 
$\left<R_{\rm Fe}\right>$ and $\left<R_{\rm Si}\right>$ are defined as average radii of 
tracer particles with $5.3 \leq T_{\rm peak}\,[{\rm GK}] \leq 5.7$ and 
$4.2 \leq T_{\rm peak}\,[{\rm GK}] \leq 4.6$, respectively. 
   \label{fig:3d-map}}
  \end{center}
\end{figure}

\begin{figure}[t]
  \begin{center}
	\includegraphics[width=7.8cm]{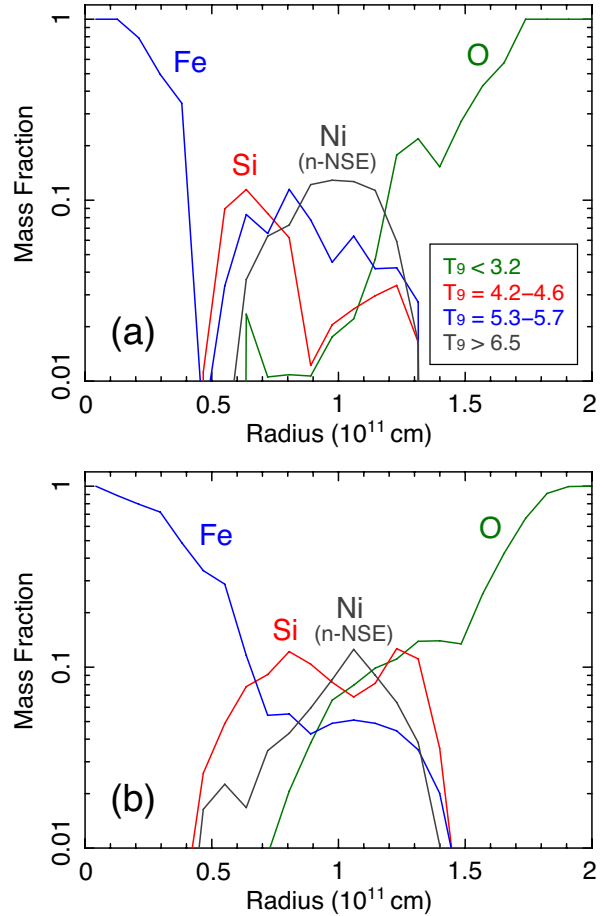}	
	\vspace{1mm}
\caption{
Radial profiles of tracer particles that have experienced peak temperatures given in the legend 
(in the unit of GK) for the Fe prominence region (a: Region A in Figure\,\ref{fig:3d-map}) 
and the nearly-opposite side (b: Region B) at $t$ = 100\,s.
The dominant element for each curve is also indicated in the panels.
   \label{fig:3d-plot}}
  \end{center}
\end{figure}

The next question is what mechanism created the Fe-rich clump and causes it to be observed now as 
a protrusion at the edge of the main SNR shell. One scenario is that the clump formed deep in 
the progenitor with sufficient density to not decelerate during the SNR evolution, and eventually 
broke out from the SNR surface \citep{Wang02,Miceli13,Tsebrenko15}. 
Another possibility is that the Fe clump rose buoyantly during the SN explosion due to some instability  
(i.e., the Fe knot has been in the outer layer since the beginning of the remnant phase). 
As a first step to assess these possibilities, we investigate whether a modern multi-dimensional SN Ia model 
naturally predicts a small-scale Fe prominence overrunning the bulk of the IMEs. 
We choose the N100 model of \cite{Seitenzahl13a} for this purpose, since it represents 
a typical SN Ia with a normal brightness, comparable with the inferred properties of {\it Tycho}'s progenitor 
\citep{Badenes06,Krause08}. We extract from the model the three-dimensional position 
of each tracer particle ($10^6$ particles in total) at $t$ = 100\,s after the initial deflagration 
ignition and the peak temperature $T_{\rm peak}$ that each particle experienced 
(see Figure\,4 of \cite{Seitenzahl13a}, for a guide). At this stage ($t$ = 100\,s), 
the SN ejecta are almost freely expanding, and non-radial velocity components are negligible. 
Therefore, an outer particle should have a higher radial velocity. 
We define $\left<R_{\rm Fe} (\theta, \phi)\right>$ as an average radius of the particles 
with $5.3 \leq T_{\rm peak}\,[{\rm GK}] \leq 5.7$
(the range constrained for the {\it Tycho}'s Fe knot) at a certain direction $(\theta, \phi)$. 
Similarly, $\left<R_{\rm Si} (\theta, \phi)\right>$ is defined as an average radius of the particles 
with $4.2 \leq T_{\rm peak}\,[{\rm GK}] \leq 4.6$, where Si is synthesized most efficiently. 
Figure\,\ref{fig:3d-map} shows the $\left<R_{\rm Fe}\right>$/$\left<R_{\rm Si}\right>$ 
distribution in the spherical coordinate system. 
We find a small region where the $\left<R_{\rm Fe}\right>$/$\left<R_{\rm Si}\right>$ value 
exceeds unity (Region A). 
Interestingly, its angular scale ($\sim$\,0.1$\pi$) is comparable with that of the {\it Tycho} Fe knot.
Virtually everywhere else, $\left<R_{\rm Fe}\right>$/$\left<R_{\rm Si}\right>$ $<$ 1, 
consistent with the observed characteristic of {\it Tycho}'s SNR.

Figure\,\ref{fig:3d-plot}(a) shows the radial distribution of the tracer particles at the Fe prominence region 
(Region~A in Figure\,\ref{fig:3d-map}) from different burning regimes.  For comparison, we show 
in Figure\,\ref{fig:3d-plot}(b) the same distribution at the nearly opposite side of the SN (Region~B). 
The presence of Fe ejecta outside the Si shell is confirmed in Region~A, 
whereas a clear layered structure is formed in the other region similarly to {\it Tycho}'s NW rim. 
In short, the state-of-the-art numerical model of delayed-detonation SNe Ia 
qualitatively reproduces the inferred Fe distribution in {\it Tycho}'s SNR. 
However, the model also predicts a relatively large radius of n-NSE burning products 
(predominantly Mn and Ni) in both regions; this is not observationally confirmed, 
at least in the Fe knot. 
Such buoyant plumes of n-NSE products are expected as a consequence of a spherically-asymmetric 
deflagration \citep{Seitenzahl13a}. In this sense, the lumpy morphology found in the {\it Suzaku} 
Ni \Ka\ image (Figure\,\ref{fig:image-suzaku}(c)) is intriguing---if real, this would be a clear signature 
of deflagration. Future study with more sensitive observations is encouraged.

\subsection{Helium Detonation?}

A possible alternative for the origin of the Fe knot is explosive He burning, 
which is thought to occur during an explosion of a sub-$M_{\rm Ch}$ white dwarf  
with a He mantle surrounding a C+O core \citep[e.g.,][]{Woosley94,Woosley11}. 
In this scenario, the first detonation ignites close to the bottom of the He mantle, 
which triggers the second detonation at the center of the white dwarf.  The detonation in 
the He-rich environment can produce almost pure $^{56}$Ni with some unburned $\alpha$-elements 
\citep[e.g.,][]{Fink07}, naturally explaining the composition of  the Fe knot; 
Knot B would then be interpreted as unburned material (see \S2.5 and \S2.6). 
However, the sub-$M_{\rm Ch}$ He detonation models tend to predict a sub-luminous SN Ia 
\citep{Fink07,Woosley11}, which is not the case for {\it Tycho}'s SNR \citep{Ruiz04,Badenes06,Krause08}.
Moreover, this explosion scenario results in the production of a substantial amount of $^{56}$Ni 
throughout the outer layers, not just one direction like in {\it Tycho}'s Fe knot \citep[e.g.,][]{Moll13}. 
The He burning also leaves behind a large amount of unburnt $\alpha$ particles, and this would 
produce a substantial thermal continuum from bremsstrahlung that is not observed in this SNR.
At this time, we cannot find any theoretical model that can perfectly reproduce 
all the observed characteristics of this SNR. If a very localized external detonation 
(in either pure He or C+O after Gyr-long gravitational deposition of $^{22}$Ne) is allowed, 
both composition and location of the Fe knot may be explained.

Finally, we remark that the presence of Fe-rich material in the outer layer is not unique to this SNR. 
A similar Fe prominence is found in the young SN Ia remnant G1.9+0.3 \citep{Borkowski13}.
Fast-moving $^{56}$Ni ejecta are also observed in the early phase of some SNe Ia, 
like SN\,2010jn \citep{Hachinger13} and SN2014J \citep{Diehl14,Isern16}. 
These observations strongly imply that the creation of Fe-rich knots is a common occurrence in SNe Ia, 
and may play an important role in the explosion mechanism itself.


\section{Conclusions}

Taking advantage of the capabilities of {\it Suzaku} and {\it Chandra}, we have performed a detailed 
spatial and spectral study of the well-known Fe-rich knot in the east region of {\it Tycho}'s SNR. 
We have shown that the ionization timescale of this knot is clearly different from that of its Si-rich 
surroundings and the rest of the Fe ejecta, identifying it as a thermodynamically and chemically 
distinct structure within the SN ejecta. 
The absence of line emission from Mn and Ni implies that the knot did not originate in the deepest 
dense layers of a $M_{\rm Ch}$ SN progenitor affected by electron capture or normal freeze-out NSE burning, 
but was instead synthesized under incomplete Si burning or $\alpha$-rich freeze out with a relatively 
low neutron excess. 
Although the composition of the Fe knot could also be the result of explosive He burning, 
this hypothesis may require a very localized external detonation; otherwise a typical He-detonation scenario 
is in conflict with the main properties of the SNR and the spectroscopy of the light echo from SN\,1572.
During the explosion, the Fe knot somehow detached from the rest of the Fe-rich material, 
and formed a protrusion at the edge of the SN ejecta, beyond most of the Si-rich material. 
We estimate a total mass for the Fe knot of  $\sim$\,$7 \times 10^{-3}$\Msun, subject to considerable uncertainty. 
The physical process responsible for the formation of the knot is unclear at this time, 
but its spatial coincidence with a previously noted break in the reverse shock traced by 
hot Fe-rich material implies that it might have played a role in the explosion mechanism itself. 
Whatever its origin is, its presence in the best observed among a handful of Type Ia SNRs with 
spatially resolved X-ray spectroscopy strongly suggests that this kind of structure might be 
relatively common in SNe Ia.

\bigskip

\acknowledgments
H.Y., C.B., H.M-R., and S.P.\ are supported by the NASA ADAP grant NNX15AM03G. 
E.B.\ is supported by the MINECO-FEDER grants AYA2013-40545 and AYA2015-63588-P.


\bigskip


\begin{thebibliography}{}
\expandafter\ifx\csname natexlab\endcsname\relax\def\natexlab#1{#1}\fi

\bibitem[{{Arnaud}(1996)}]{Arnaud96}
{Arnaud}, K.~A. 1996, in Astronomical Society of the Pacific Conference Series,
  Vol. 101, Astronomical Data Analysis Software and Systems V, ed. G.~H.
  {Jacoby} \& J.~{Barnes}, 17

\bibitem[{{Asplund} {et~al.}(2009){Asplund}, {Grevesse}, {Sauval}, \&
  {Scott}}]{Asplund09}
{Asplund}, M., {Grevesse}, N., {Sauval}, A.~J., \& {Scott}, P. 2009, \araa, 47,
  481

\bibitem[{{Badenes} {et~al.}(2006){Badenes}, {Borkowski}, {Hughes}, {Hwang}, \&
  {Bravo}}]{Badenes06}
{Badenes}, C., {Borkowski}, K.~J., {Hughes}, J.~P., {Hwang}, U., \& {Bravo}, E.
  2006, \apj, 645, 1373

\bibitem[{{Badenes} {et~al.}(2003){Badenes}, {Bravo}, {Borkowski}, \&
  {Dom{\'{\i}}nguez}}]{Badenes03}
{Badenes}, C., {Bravo}, E., {Borkowski}, K.~J., \& {Dom{\'{\i}}nguez}, I. 2003,
  \apj, 593, 358

\bibitem[{{Badenes} {et~al.}(2008){Badenes}, {Bravo}, \& {Hughes}}]{Badenes08b}
{Badenes}, C., {Bravo}, E., \& {Hughes}, J.~P. 2008, \apjl, 680, L33

\bibitem[{{Badenes} {et~al.}(2007){Badenes}, {Hughes}, {Bravo}, \&
  {Langer}}]{Badenes07}
{Badenes}, C., {Hughes}, J.~P., {Bravo}, E., \& {Langer}, N. 2007, \apj, 662,
  472

\bibitem[{{Borkowski} {et~al.}(2013){Borkowski}, {Reynolds}, {Hwang}, {Green},
  {Petre}, {Krishnamurthy}, \& {Willett}}]{Borkowski13}
{Borkowski}, K.~J., {Reynolds}, S.~P., {Hwang}, U., {et~al.} 2013, \apjl, 771,
  L9

\bibitem[{{Bravo} \& {Mart{\'{\i}}nez-Pinedo}(2012)}]{Bravo12}
{Bravo}, E., \& {Mart{\'{\i}}nez-Pinedo}, G. 2012, \prc, 85, 055805

\bibitem[{{Cassam-Chena{\"i}} {et~al.}(2007){Cassam-Chena{\"i}}, {Hughes},
  {Ballet}, \& {Decourchelle}}]{Cassam07}
{Cassam-Chena{\"i}}, G., {Hughes}, J.~P., {Ballet}, J., \& {Decourchelle}, A.
  2007, \apj, 665, 315

\bibitem[{{Decourchelle} {et~al.}(2001){Decourchelle}, {Sauvageot}, {Audard},
  {Aschenbach}, {Sembay}, {Rothenflug}, {Ballet}, {Stadlbauer}, \&
  {West}}]{Decourchelle01}
{Decourchelle}, A., {Sauvageot}, J.~L., {Audard}, M., {et~al.} 2001, \aap, 365,
  L218

\bibitem[{{Diehl} {et~al.}(2014){Diehl}, {Siegert}, {Hillebrandt}, {Grebenev},
  {Greiner}, {Krause}, {Kromer}, {Maeda}, {R{\"o}pke}, \&
  {Taubenberger}}]{Diehl14}
{Diehl}, R., {Siegert}, T., {Hillebrandt}, W., {et~al.} 2014, Science, 345,
  1162

\bibitem[{{Eriksen} {et~al.}(2011){Eriksen}, {Hughes}, {Badenes}, {Fesen},
  {Ghavamian}, {Moffett}, {Plucinksy}, {Rakowski}, {Reynoso}, \&
  {Slane}}]{Eriksen11}
{Eriksen}, K.~A., {Hughes}, J.~P., {Badenes}, C., {et~al.} 2011, \apj, 728, L28

\bibitem[{{Fink} {et~al.}(2007){Fink}, {Hillebrandt}, \& {R{\"o}pke}}]{Fink07}
{Fink}, M., {Hillebrandt}, W., \& {R{\"o}pke}, F.~K. 2007, \aap, 476, 1133

\bibitem[{{Hachinger} {et~al.}(2013){Hachinger}, {Mazzali}, {Sullivan},
  {Ellis}, {Maguire}, {Gal-Yam}, {Howell}, {Nugent}, {Baron}, {Cooke},
  {Arcavi}, {Bersier}, {Dilday}, {James}, {Kasliwal}, {Kulkarni}, {Ofek},
  {Laher}, {Parrent}, {Surace}, {Yaron}, \& {Walker}}]{Hachinger13}
{Hachinger}, S., {Mazzali}, P.~A., {Sullivan}, M., {et~al.} 2013, \mnras, 429,
  2228

\bibitem[{{Hayato} {et~al.}(2010){Hayato}, {Yamaguchi}, {Tamagawa}, {Katsuda},
  {Hwang}, {Hughes}, {Ozawa}, {Bamba}, {Kinugasa}, {Terada}, {Furuzawa},
  {Kunieda}, \& {Makishima}}]{Hayato10}
{Hayato}, A., {Yamaguchi}, H., {Tamagawa}, T., {et~al.} 2010, \apj, 725, 894

\bibitem[{{Hwang} \& {Gotthelf}(1997)}]{Hwang97}
{Hwang}, U., \& {Gotthelf}, E.~V. 1997, \apj, 475, 665

\bibitem[{{Hwang} {et~al.}(1998){Hwang}, {Hughes}, \& {Petre}}]{Hwang98}
{Hwang}, U., {Hughes}, J.~P., \& {Petre}, R. 1998, \apj, 497, 833

\bibitem[{{Hwang} \& {Laming}(2012)}]{Hwang12}
{Hwang}, U., \& {Laming}, J.~M. 2012, \apj, 746, 130

\bibitem[{{Isern} {et~al.}(2016){Isern}, {Jean}, {Bravo}, {Kn{\"o}dlseder},
  {Lebrun}, {Churazov}, {Sunyaev}, {Domingo}, {Badenes}, {Hartmann},
  {Hoeflich}, {Renaud}, {Soldi}, {Elias-Rosa}, {Hernanz}, {Dom{\'{\i}}nguez},
  {Garc{\'{\i}}a-Senz}, {Lichti}, {Vedrenne}, \& {Von Ballmoos}}]{Isern16}
{Isern}, J., {Jean}, P., {Bravo}, E., {et~al.} 2016, \aap, 588, A67

\bibitem[{{Iwamoto} {et~al.}(1999){Iwamoto}, {Brachwitz}, {Nomoto},
  {Kishimoto}, {Umeda}, {Hix}, \& {Thielemann}}]{Iwamoto99}
{Iwamoto}, K., {Brachwitz}, F., {Nomoto}, K., {et~al.} 1999, \apjs, 125, 439

\bibitem[{{Krause} {et~al.}(2008){Krause}, {Tanaka}, {Usuda}, {Hattori},
  {Goto}, {Birkmann}, \& {Nomoto}}]{Krause08}
{Krause}, O., {Tanaka}, M., {Usuda}, T., {et~al.} 2008, \nat, 456, 617

\bibitem[{{Maoz} {et~al.}(2014){Maoz}, {Mannucci}, \& {Nelemans}}]{Maoz14}
{Maoz}, D., {Mannucci}, F., \& {Nelemans}, G. 2014, \araa, 52, 107

\bibitem[{{Mart{\'{\i}}nez-Rodr{\'{\i}}guez}
  {et~al.}(2016){Mart{\'{\i}}nez-Rodr{\'{\i}}guez}, {Piro}, {Schwab}, \&
  {Badenes}}]{Martinez16}
{Mart{\'{\i}}nez-Rodr{\'{\i}}guez}, H., {Piro}, A.~L., {Schwab}, J., \&
  {Badenes}, C. 2016, \apj, 825, 57

\bibitem[{{Mazzali} {et~al.}(2007){Mazzali}, {R{\"o}pke}, {Benetti}, \&
  {Hillebrandt}}]{Mazzali07}
{Mazzali}, P.~A., {R{\"o}pke}, F.~K., {Benetti}, S., \& {Hillebrandt}, W. 2007,
  Science, 315, 825

\bibitem[{{Miceli} {et~al.}(2013){Miceli}, {Orlando}, {Reale}, {Bocchino}, \&
  {Peres}}]{Miceli13}
{Miceli}, M., {Orlando}, S., {Reale}, F., {Bocchino}, F., \& {Peres}, G. 2013,
  \mnras, 430, 2864

\bibitem[{{Miceli} {et~al.}(2015){Miceli}, {Sciortino}, {Troja}, \&
  {Orlando}}]{Miceli15}
{Miceli}, M., {Sciortino}, S., {Troja}, E., \& {Orlando}, S. 2015, \apj, 805,
  120

\bibitem[{{Moll} \& {Woosley}(2013)}]{Moll13}
{Moll}, R., \& {Woosley}, S.~E. 2013, \apj, 774, 137

\bibitem[{{Ozawa} {et~al.}(2009){Ozawa}, {Uchiyama}, {Matsumoto}, {Nakajima},
  {Koyama}, {Tsuru}, {Uchino}, {Uchida}, {Hayashida}, {Tsunemi}, {Mori},
  {Bamba}, {Ozaki}, {Dotani}, {Kohmura}, {Ishisaki}, {Murakami}, {Kato},
  {Kitazono}, {Kimura}, {Ogawa}, {Kawai}, {Mori}, {Prigozhin}, {Kissel},
  {Miller}, {Lamarr}, \& {Bautz}}]{Ozawa09a}
{Ozawa}, M., {Uchiyama}, H., {Matsumoto}, H., {et~al.} 2009, \pasj, 61, 1

\bibitem[{{Park} {et~al.}(2007){Park}, {Hughes}, {Slane}, {Burrows},
  {Gaensler}, \& {Ghavamian}}]{Park07}
{Park}, S., {Hughes}, J.~P., {Slane}, P.~O., {et~al.} 2007, \apjl, 670, L121

\bibitem[{{Park} {et~al.}(2013){Park}, {Badenes}, {Mori}, {Kaida}, {Bravo},
  {Schenck}, {Eriksen}, {Hughes}, {Slane}, {Burrows}, \& {Lee}}]{Park13}
{Park}, S., {Badenes}, C., {Mori}, K., {et~al.} 2013, \apjl, 767, L10

\bibitem[{{Piro} \& {Bildsten}(2008)}]{Piro08}
{Piro}, A.~L., \& {Bildsten}, L. 2008, \apj, 673, 1009

\bibitem[{{Ruiz-Lapuente}(2004)}]{Ruiz04}
{Ruiz-Lapuente}, P. 2004, \apj, 612, 357

\bibitem[{{Sato} \& {Hughes}(2016)}]{Sato16}
{Sato}, T., \& {Hughes}, J.~P. 2016, ArXiv e-prints, arXiv:1605.09059

\bibitem[{{Seitenzahl} {et~al.}(2013{\natexlab{a}}){Seitenzahl}, {Cescutti},
  {R{\"o}pke}, {Ruiter}, \& {Pakmor}}]{Seitenzahl13b}
{Seitenzahl}, I.~R., {Cescutti}, G., {R{\"o}pke}, F.~K., {Ruiter}, A.~J., \&
  {Pakmor}, R. 2013{\natexlab{a}}, \aap, 559, L5

\bibitem[{{Seitenzahl} {et~al.}(2013{\natexlab{b}}){Seitenzahl},
  {Ciaraldi-Schoolmann}, {R{\"o}pke}, {Fink}, {Hillebrandt}, {Kromer},
  {Pakmor}, {Ruiter}, {Sim}, \& {Taubenberger}}]{Seitenzahl13a}
{Seitenzahl}, I.~R., {Ciaraldi-Schoolmann}, F., {R{\"o}pke}, F.~K., {et~al.}
  2013{\natexlab{b}}, \mnras, 429, 1156

\bibitem[{{Serlemitsos} {et~al.}(2007){Serlemitsos}, {Soong}, {Chan},
  {Okajima}, {Lehan}, {Maeda}, {Itoh}, {Mori}, {Iizuka}, {Itoh}, {Inoue},
  {Okada}, {Yokoyama}, {Itoh}, {Ebara}, {Nakamura}, {Suzuki}, {Ishida},
  {Hayakawa}, {Inoue}, {Okuma}, {Kubota}, {Suzuki}, {Osawa}, {Yamashita},
  {Kunieda}, {Tawara}, {Ogasaka}, {Furuzawa}, {Tamura}, {Shibata}, {Haba},
  {Naitou}, \& {Misaki}}]{Serlemitsos07}
{Serlemitsos}, P.~J., {Soong}, Y., {Chan}, K.-W., {et~al.} 2007, \pasj, 59, 9


\bibitem[{{Tamagawa} {et~al.}(2009){Tamagawa}, {Hayato}, {Nakamura}, {Terada},
  {Bamba}, {Hiraga}, {Hughes}, {Hwang}, {Kataoka}, {Kinugasa}, {Kunieda},
  {Tanaka}, {Tsunemi}, {Ueno}, {Holt}, {Kokubun}, {Miyata}, {Szymkowiak},
  {Takahashi}, {Tamura}, {Ueno}, \& {Makishima}}]{Tamagawa09}
{Tamagawa}, T., {Hayato}, A., {Nakamura}, S., {et~al.} 2009, \pasj, 61, S167

\bibitem[{{Tanaka} {et~al.}(2011){Tanaka}, {Mazzali}, {Stanishev}, {Maurer},
  {Kerzendorf}, \& {Nomoto}}]{Tanaka10}
{Tanaka}, M., {Mazzali}, P.~A., {Stanishev}, V., {et~al.} 2011, \mnras, 410,
  1725

\bibitem[{{Thielemann} {et~al.}(1986){Thielemann}, {Nomoto}, \&
  {Yokoi}}]{Thielemann86}
{Thielemann}, F.-K., {Nomoto}, K., \& {Yokoi}, K. 1986, \aap, 158, 17

\bibitem[{{Tian} \& {Leahy}(2011)}]{Tian11}
{Tian}, W.~W., \& {Leahy}, D.~A. 2011, \apjl, 729, L15

\bibitem[{{Timmes} {et~al.}(2003){Timmes}, {Brown}, \& {Truran}}]{Timmes03}
{Timmes}, F.~X., {Brown}, E.~F., \& {Truran}, J.~W. 2003, \apjl, 590, L83

\bibitem[{{Tsebrenko} \& {Soker}(2015)}]{Tsebrenko15}
{Tsebrenko}, D., \& {Soker}, N. 2015, \mnras, 453, 166

\bibitem[{{Vancura} {et~al.}(1995){Vancura}, {Gorenstein}, \&
  {Hughes}}]{Vancura95}
{Vancura}, O., {Gorenstein}, P., \& {Hughes}, J.~P. 1995, \apj, 441, 680

\bibitem[{{Wang} \& {Chevalier}(2002)}]{Wang02}
{Wang}, C.-Y., \& {Chevalier}, R.~A. 2002, \apj, 574, 155

\bibitem[{{Warren} {et~al.}(2005){Warren}, {Hughes}, {Badenes}, {Ghavamian},
  {McKee}, {Moffett}, {Plucinsky}, {Rakowski}, {Reynoso}, \&
  {Slane}}]{Warren05}
{Warren}, J.~S., {Hughes}, J.~P., {Badenes}, C., {et~al.} 2005, \apj, 634, 376

\bibitem[{{Williams} {et~al.}(2013){Williams}, {Borkowski}, {Ghavamian},
  {Hewitt}, {Mao}, {Petre}, {Reynolds}, \& {Blondin}}]{Williams13}
{Williams}, B.~J., {Borkowski}, K.~J., {Ghavamian}, P., {et~al.} 2013, \apj,
  770, 129

\bibitem[{{Wilms} {et~al.}(2000){Wilms}, {Allen}, \& {McCray}}]{Wilms00}
{Wilms}, J., {Allen}, A., \& {McCray}, R. 2000, \apj, 542, 914

\bibitem[{{Woosley} \& {Kasen}(2011)}]{Woosley11}
{Woosley}, S.~E., \& {Kasen}, D. 2011, \apj, 734, 38

\bibitem[{{Woosley} \& {Weaver}(1994)}]{Woosley94}
{Woosley}, S.~E., \& {Weaver}, T.~A. 1994, \apj, 423, 371

\bibitem[{{Yamaguchi} {et~al.}(2008){Yamaguchi}, {Koyama}, {Katsuda},
  {Nakajima}, {Hughes}, {Bamba}, {Hiraga}, {Mori}, {Ozaki}, \&
  {Tsuru}}]{Yamaguchi08b}
{Yamaguchi}, H., {Koyama}, K., {Katsuda}, S., {et~al.} 2008, \pasj, 60, S141

\bibitem[{{Yamaguchi} {et~al.}(2014{\natexlab{a}}){Yamaguchi}, {Badenes},
  {Petre}, {Nakano}, {Castro}, {Enoto}, {Hiraga}, {Hughes}, {Maeda},
  {Nobukawa}, {Safi-Harb}, {Slane}, {Smith}, \& {Uchida}}]{Yamaguchi14b}
{Yamaguchi}, H., {Badenes}, C., {Petre}, R., {et~al.} 2014{\natexlab{a}},
  \apjl, 785, L27

\bibitem[{{Yamaguchi} {et~al.}(2014{\natexlab{b}}){Yamaguchi}, {Eriksen},
  {Badenes}, {Hughes}, {Brickhouse}, {Foster}, {Patnaude}, {Petre}, {Slane}, \&
  {Smith}}]{Yamaguchi14a}
{Yamaguchi}, H., {Eriksen}, K.~A., {Badenes}, C., {et~al.} 2014{\natexlab{b}},
  \apj, 780, 136

\bibitem[{{Yamaguchi} {et~al.}(2015){Yamaguchi}, {Badenes}, {Foster}, {Bravo},
  {Williams}, {Maeda}, {Nobukawa}, {Eriksen}, {Brickhouse}, {Petre}, \&
  {Koyama}}]{Yamaguchi15}
{Yamaguchi}, H., {Badenes}, C., {Foster}, A.~R., {et~al.} 2015, \apjl, 801, L31

\bibitem[{{Yang} {et~al.}(2013){Yang}, {Tsunemi}, {Lu}, {Li}, {Xiang}, {Xiao},
  \& {Zhong}}]{Yang13}
{Yang}, X.~J., {Tsunemi}, H., {Lu}, F.~J., {et~al.} 2013, \apj, 766, 44

\bibitem[{{Yasumi} {et~al.}(2014){Yasumi}, {Nobukawa}, {Nakashima}, {Uchida},
  {Sugawara}, {Tsuru}, {Tanaka}, \& {Koyama}}]{Yasumi14}
{Yasumi}, M., {Nobukawa}, M., {Nakashima}, S., {et~al.} 2014, \pasj, 66, 68

\end{thebibliography}
\end{document}